\documentclass[12pt]{article}

\def\newpic#1{%
   \def\emline##1##2##3##4##5##6{%
      \put(##1,##2){\special{em:point #1##3}}%
      \put(##4,##5){\special{em:point #1##6}}%
      \special{em:line #1##3,#1##6}}}
\newpic{}

\newif\ifnotesw\noteswtrue
\newcommand{\comm}[1]{\ifnotesw $\blacktriangleright$\ {\sf #1}\ 
  $\blacktriangleleft$ \fi}
\noteswfalse	

\usepackage{amssymb} 
   
\newcommand{\hide}[1]{}

\def\afterthmseparator{.}
\makeatletter
\renewcommand{\@begintheorem}[2]{\trivlist
      \item[\hskip \labelsep{\bf #1\ #2\unskip\afterthmseparator}]\em}
\renewcommand{\@opargbegintheorem}[3]{\trivlist
      \item[\hskip \labelsep{\bf #1\ #2\ (#3)\unskip\afterthmseparator}]\em}
\makeatother
\newtheorem{theorem}{Theorem}[section]
\newtheorem{lemma}[theorem]{Lemma}
\newtheorem{proposition}[theorem]{Proposition}
\newtheorem{corollary}[theorem]{Corollary}

\newtheorem{definition}[theorem]{Definition}
\newtheorem{convention}[theorem]{Convention}
\newtheorem{assumption}[theorem]{Assumption}

\newcommand{\bull}{\mbox{$\;\;\;$\vrule height .9ex width .8ex depth -.1ex}}
\newcommand{\qed}{$\;\;\;\Box$}
\newenvironment{proof}{\par\smallbreak\noindent{\bf Proof.~}}
{\unskip\nobreak\hfill \bull \par\medbreak}

{\qed \par\smallbreak}
\newcommand{\noproof}{\unskip\nobreak\hfill \bull}
\newcounter{claim}[theorem]
\renewcommand{\theclaim}{\thetheorem.\arabic{claim}}
{\par\smallskip\par}
\newcommand{\subcase}[2]{\smallskip\par\noindent{\it Subcase #1:\/ #2}}
\newcommand{\case}[2]%
{\par\medskip\par\noindent{\it Case #1:\/ #2}\par\smallskip\par\noindent}
\newenvironment{bfenumerate}%
{\begin{enumerate}}%
{\end{enumerate}}

\newcommand{\assfont}{\rm}

\newcommand{\refeq}[1]{(\ref{#1})}

\newcommand{\setdef}[2]{\left\{ \hspace{0.5mm} #1 :
\hspace{0.5mm} #2 \right\}}

\newcommand{\function}[2]{:#1 \rightarrow #2}

\newcommand{\diam}[1]{\mathit{diam}\,(#1)}

\newcommand{\cd}[2]{{\scriptstyle c}D^{#1}(#2)}
\newcommand{\game}{\mathrm{Ehr}}   
\newcommand{\EF}{Ehrenfeucht-Fra\"\i{}ss\'{e}}

\newcommand{\WLe}{Weisfeiler-Lehman}

\newcommand{\tc}[1]{\mbox{\rm TC$^{#1}$}}

\newcommand{\ac}[1]{\mbox{\rm AC$^{#1}$}}
\newcommand{\naturals}{{\mathbb N}}
\newcommand{\calS}{{\mathcal S}}
\newcommand{\calP}{{\mathcal P}}
\newcommand{\calQ}{{\mathcal Q}}
\newcommand{\lay}{\mathsf{L}}

\title{Planar Graphs:
Logical Complexity and\\
Parallel Isomorphism Tests}

\author{Oleg Verbitsky%
\thanks{Supported by an Alexander von Humboldt fellowship.}\\[4mm]
Institut f\"ur Informatik\\
Humboldt Universit\"at zu Berlin, D-10099 Berlin}

\date{8 July 2006}

\begin{document}

\sloppy

\maketitle

\begin{abstract}
We prove that every triconnected planar graph is definable by a first order
sentence that uses at most 15 variables and has quantifier depth at most $11\,\log_2 n+43$.
As a consequence, a canonic form of such graphs is computable in \ac1\
by the 14-dimensional \WLe\/ algorithm. This provides another way to show that
the planar graph isomorphism is solvable in \ac1.
\end{abstract}

\section{Introduction}

Let $\Phi$ be a first order sentence about graphs in terms of
the adjacency and the equality relations. We say that $\Phi$ \emph{distinguishes}
a graph $G$ from a graph $H$ if $\Phi$ is true on $G$ but false on $H$.
We say that $\Phi$ \emph{defines} $G$ if it distinguishes $G$ from every $H$
non-isomorphic to $G$.
The \emph{logical depth} of a graph $G$, denoted by $D(G)$, is the minimum
quantifier depth of a $\Phi$ defining $G$.

The \emph{$k$-variable logic} consists of those first order sentences which
use at most $k$ variables (each of the $k$ variables can occur a number of times).
The \emph{logical width} of a graph $G$, denoted by $W(G)$, is the minimum $k$
such that $G$ is definable by a $\Phi$ in the $k$-variable logic.
If $k\ge W(G)$, let $D^k(G)$ denote the logical depth of $G$ in the $k$-variable
logic. Similarly, for non-isomorphic graphs $G$ and $H$ we 
let $D^k(G,H)$ denote the minimum quantifier depth of a $k$-variable
sentence $\Phi$ distinguishing $G$ from $H$.

The latter parameter is relevant to the Graph Isomorphism problem,
namely, to the \emph{$k$-dimensional \WLe\ algorithm} (see \cite{Bab,CFI} for
the description and history).
Cai, F\"urer, and Immerman \cite{CFI} proved that, if $k > W(G)$, then 
the output of this algorithm is correct for all input pairs $(G,H)$.
The above condition on the dimension $k$ is necessary if we consider the width
of $G$ in the logic with \emph{counting quantifiers}.
Furthermore, Cai, F\"urer, and Immerman constructed examples of $G$ for which
the latter parameter is linear in the number of vertices.

Note that the $k$-dimensional \WLe\ algorithm is polynomial-time only
if the dimension is constant. Thus, the algorithm can be successful only
for classes of graphs whose width in the logic with counting quantifiers
is bounded by a constant. Cai, F\"urer, and Immerman asked if this is
the case for planar graphs. An affirmative answer was given by Grohe~\cite{Gro1}.

In \cite{GVe} we extended the approach to Graph Isomorphism suggested in \cite{CFI} 
by emphasizing not only on the dimension but also
on the number of rounds performed by the \WLe\ algorithm.
Namely, the logarithmic-round $k$-dimensional \WLe\ algorithm is implementable
in \tc1 and its count-free version even in \ac1.
We applied this fact to show that the isomorphism problem for graphs
of bounded treewidth is in \tc1 (earlier Grohe and Marino \cite{GMa}
proved that such graphs have bounded width in the logic with counting).

By the framework worked out in \cite{GVe}, to put the isomorphism problem
for a class of graphs $C$ in \ac1, it suffices to prove that, for a constant $k$,
we have $D^k(G,G')=O(\log n)$ for all $G$ and $G'$ in $C$.
We now apply this approach to planar graphs.

\begin{theorem}\label{thm:dist3}
Let $G$ and $G'$ be non-isomorphic triconnected planar graphs
and $G$ have $n$ vertices. Then $D^{15}(G,G')<11\,\log_2 n+43$.
\end{theorem}

Within the framework of \cite{GVe}, Theorem \ref{thm:dist3}
allows us to reprove a result of Miller and Reif~\cite{MRe}.

\begin{corollary}
The isomorphism problem for triconnected planar graphs is solvable in~\ac1.
\end{corollary}

It should be stressed that our algorithm is different: 
it is essentially the logarithmic-round 14-dimensional
\WLe\/ algorithm (see \cite{GVe} for details).

With not so much extra work, we are able to strengthen Theorem~\ref{thm:dist3}.

\begin{theorem}\label{thm:def3}
For a triconnected planar graph $G$ on $n$ vertices we have $D^{15}(G)<11\,\log_2 n+43$.
\end{theorem}

In the framework of \cite{GVe}, this means that an appropriate modification of 
the logarithmic-round 14-dimensional \WLe\/ algorithm computes a canonic form of
a triconnected planar input graph, putting this problem in the class \ac1. 
Miller and Reif \cite{MRe} show that the canonization of planar graphs \ac1-reduces
to the triconnected case. Using this reduction, we hence obtain a new \ac1-algorithm
for the planar graph isomorphism problem.

\begin{corollary}
The canonization problem for planar graphs is solvable in~\ac1.
\end{corollary} 

Theorem \ref{thm:def3} is also a contribution in a recent line of research
\cite{KPSV,PSV,PVV}
devoted to a general study of the logical depth $D(G)$ as a mysterious graph
invariant.

\section{Basic definitions and notation}

Throughout the paper $\log n$ denotes the binary logarithm.
Unless stated otherwise, $n$ will denote the number of vertices in a graph $G$.

\subsection{Graphs}\label{ss:graphs}

The vertex set of a graph $G$ is denoted by $V(G)$. Let $\ell\in\naturals$.
A sequence of pairwise distinct vertices $u=w_0,\ldots,w_{\ell}=v$ 
where every two successive vertices
are adjacent is called a \emph{$u$-$v$-path}
(if $u=v$ and $\ell\ge2$, we have a \emph{cycle}). 
The $u$ and $v$ are the two \emph{endpoints};
any $w_i$, $0\le i<\ell$, is an \emph{inner point}. The number $\ell$ is referred to
as the \emph{length} of the path. If we denote such a path by $P$, then its segment
$w_i,\ldots,w_j$ will be denoted by $P[w_i,w_j]$.
Given $X\subset V(G)$, we say that a path $P$ \emph{avoids} $X$ if it has
no inner point in~$X$.

A graph is \emph{connected} if a $u$-$v$-path exists for every two vertices $u$ and $v$.
A graph is \emph{$k$-connected} if it has at least $k+1$ vertices and 
remains connected after removal of any $k-1$ vertices.

The distance between vertices $u$ and $v$ in a graph $G$ is 
defined to be the minimum length of a $u$-$v$-path and denoted by $d(u,v)$.
If $u$ and $v$ are in different connected components, we set $d(u,v)=\infty$.
A $u$-$v$-path having the smallest length $d(u,v)$ will be sometimes referred to
as a \emph{geodesic} between these vertices.
The set $\Gamma(v)=\setdef{u}{d(u,v)=1}$ is called the \emph{neighborhood}
of a vertex $v$ in $G$.

Let $X\subset V(G)$. The subgraph induced by $G$ on $X$ is denoted by $G[X]$.
We denote $G\setminus X=G[V(G)\setminus X]$, which is the result of removal of all
vertices in $X$ from $G$. 
If a single vertex $v$ is removed, we write $G-v=G\setminus\{v\}$.

Suppose that $G$ is connected. A vertex $v$ is a \emph{cutpoint} of $G$
if $G-v$ is disconnected. An edge $e$ of $G$ is called a \emph{bridge} if
$e$ belongs to no cycle. A \emph{block} is a maximal subgraph of $G$ with no cutpoint.
Thus, every block is either a maximal biconnected subgraph or a bridge.
Every two blocks share at most one cutpoint. If every cutpoint belongs to at most
two blocks, we will say that $G$ has \emph{simple cut-block relation}.
Suppose that $G$ has this property and consider a graph $B(G)$ whose vertices are
the blocks of $G$ and two blocks are adjacent if they share a cutpoint.
It is not hard to see that $B(G)$ is a tree and we will call it the 
\emph{block-tree} of~$G$.

A \emph{sphere graph} is a graph drawn in a sphere with no edge crossing
(here and below we refer the reader to \cite{Die} for a systematic account). 
A \emph{spherical embedding} of a graph $G$ is an isomorphism from $G$ to a sphere graph
$\tilde G$. We call $G$ \emph{planar} if it has a spherical embedding (which is
equivalent to the condition that $G$ has a plane embedding). Two spherical embeddings
$\sigma\function G{\tilde G}$ and $\tau\function G{\hat G}$ are equivalent if
the isomorphism $\tau\circ\sigma^{-1}$ is induced by a homeomorphism of a sphere
taking $\tilde G$ onto $\hat G$.
The Whitney theorem says that all spherical embeddings of a triconnected planar
graph $G$ are equivalent.

\subsection{Logic}

Let $\Phi$ be a first order sentence about a graph in the language
of the adjacency and the equality relations. We say that $\Phi$
\emph{distinguishes} a graph $G$ from a graph $H$ if $\Phi$ is true
on $G$ but false on $H$. We say that $\Phi$
\emph{defines} $G$ if $\Phi$ is true on $G$ and false on any graph
non-isomorphic to $G$. The quantifier rank of $\Phi$ is the maximum number of nested
quantifiers in $\Phi$. The \emph{logical depth} of a graph $G$, denoted by $D(G)$, 
is the minimum quantifier depth of $\Phi$ defining $G$.

The \emph{$k$-variable logic} is the fragment of first order logic where
usage of only $k$ variables is allowed. 
If we restrict defining sentences to the $k$-variable logic, this variant
of the logical depth of $G$ is denoted by $D^k(G)$.
We have 
\begin{equation}\label{eq:ddd}
D^k(G)=\max\setdef{D^k(G,H)}{H\not\cong G},
\end{equation}
where $D^k(G,H)$ denotes the minimum quantifier depth of a $k$-variable sentence
distinguishing $G$ from $H$.
This equality easily follows from the fact that, for each $r$, 
there are only finitely many pairwise inequivalent first order sentences
about graphs of quantifier depth at most $r$. It is assumed that
$D^k(G)=\infty$ (resp.\ $D^k(G,H)=\infty$) if the $k$-variable logic is too
weak to define $G$ (resp.\ to distinguish $G$ from $H$).

Furthermore, let $\cd kG$ (resp.\ $\cd k{G,H}$) denote the variant of $D^k(G)$
(resp.\ $D^k(G,H)$) for the first order logic
with \emph{counting quantifiers} where we allow expressions of the type
$\exists^m\Psi$ to say that there are at least $m$ vertices with property $\Psi$
(such a quantifier contributes 1 in the quantifier depth irrespective of $m$).
Similarly to \refeq{eq:ddd} we have
\begin{equation}\label{eq:dddc}
\cd kG=\max\setdef{\cd k{G,H}}{H\not\cong G}.
\end{equation}

\subsection{Games}\label{s:ehr}

Let $G$ and $G'$ be graphs with disjoint vertex sets.
The \emph{$r$-round $k$-pebble \EF\/ game on $G$ and $G$},
denoted by $\game_r^k(G,G')$, is played by
two players, Spoiler and Duplicator, with $k$ pairwise distinct
pebbles $p_1,\ldots,p_k$, each given in duplicate. Spoiler starts the game.
A {\em round\/} consists of a move of Spoiler followed by a move of
Duplicator. At each move Spoiler takes a pebble, say $p_i$, selects one of
the graphs $G$ or $G'$, and places $p_i$ on a vertex of this graph.
In response Duplicator should place the other copy of $p_i$ on a vertex
of the other graph. It is allowed to remove previously placed pebbles
to another vertex and place more than one pebble on the same vertex.

After each round of the game, for $1\le i\le k$ let $x_i$ (resp.\ $x'_i$)
denote the vertex of $G$ (resp.\ $G'$) occupied by $p_i$, irrespectively
of who of the players placed the pebble on this vertex. If $p_i$ is
off the board at this moment, $x_i$ and $x'_i$ are undefined.
If after every of $r$ rounds the component-wise correspondence $(x_1,\ldots,x_k)$ to
$(x'_1,\ldots,x'_k)$ is a partial isomorphism from $G$ to $G'$, this is
a win for Duplicator;  Otherwise the winner is Spoiler.

Let $\bar v=(v_1,\ldots,v_m)$ and $\bar v'=(v'_1,\ldots,v'_m)$ be sequences of
vertices in, respectively, $G$ and $G'$ and let $m\le k$.
We write $\game_r^k(G,\bar v,G',\bar v')$ to denote the game that begins from
the position where, for every $i\le m$, the vertices $v_i$ and $v'_i$ are already pebbled
by~$p_i$.

In the \emph{counting version} of the game,
the rules of $\game_r^k(G,G')$ are modified as follows.
A round now consists of two acts. First, Spoiler specifies a set of vertices
$A$ in one of the graphs. Duplicator responds with a set of vertices $B$
in the other graph so that $|B|=|A|$. Second, Spoiler places a pebble $p_i$
on a vertex $b\in B$. In response Duplicator has to place the other copy
of $p_i$ on a vertex $a\in A$.
We will say that Spoiler makes a \emph{composite move}.

\begin{proposition}\label{prop:game}{\bf (Immerman, Poizat, see \cite[Theorem 6.10]{Imm})}
\begin{bfenumerate}
\item
$D^k(G,G')$ equals the minimum $r$ such that Spoiler has a winning
strategy in $\game_r^k(G,G')$.
\item
$\cd k{G,G'}$ equals the minimum $r$ such that Spoiler has a winning
strategy in the counting version of $\game_r^k(G,G')$.
\end{bfenumerate}
\end{proposition}

All the above definitions and statements have a perfect sense
for any kind of structures. Say,
in Section \ref{s:global} we deal with structures having
one binary and one ternary relations. The notion of a partial
isomorphism for such structures should be understood appropriately.

For our convenience, everywhere below it is assumed that vertex names correspond to pebbling;
for example, vertices $v\in V(G)$ and $v'\in V(G')$ are always under the same pebbles.
We will refer to Spoiler as \emph{him} and to Duplicator as \emph{her}. Furthermore,
we will write \emph{Spoiler wins} with meaning that \emph{Spoiler has a strategy that wins
against any Duplicator's strategy.}

The following fact is based on a well-known trick which is used throughout
the paper in many variations.

\begin{lemma}{\bf(Halving Strategy)}\label{lem:halving}
Consider the \EF\/ game on graphs $G$ and $G'$. 
Let $u,v\in V(G)$, $u',v'\in V(G')$ and suppose that
$d(u,v)\ne d(u',v')$ and $d(u,v)\ne\infty$
(in particular, it is possible that $d(u',v')=\infty$).
Then Spoiler wins $\game^3_r(G,u,v,G',u',v')$ where $r=\lceil\log d(u,v)\rceil$.
\end{lemma}

\begin{proof}
Without loss of generality, assume $d(u,v) < d(u',v')$. Spoiler uses a \emph{halving strategy}
(see \cite[Theorem 2.1.2]{Spe} for a detailed account). 
In the first round he pebbles a vertex $w$ on the midway between $u$ and $v$. 
Note that $d(u,w) < d(u',w')$ or $d(w,v) < d(w',v')$.
Spoiler selects a pair for which the inequality is true and repeats the same trick
for this pair, reusing the pebble from the remaining vertex. Eventually Spoiler forces
pebbling vertices $a,b\in V(G)$ and $a',b'\in V(G')$ so that $d(a,b)=1$ while $d(a,b)>1$,
which means his win.
\end{proof}

Let us state a simple consequence.

\begin{lemma}\label{lem:drespect}
Duplicator is forced to respect the graph metric $d$: Once vertices 
$u,v\in V(G)$ and $u',v'\in V(G')$ are pebbled so that $d(u,v)\ne d(u',v')$,
Spoiler wins operating with 3 pebbles in at most $\lceil\log n\rceil+1$ extra moves,
where $n$ denotes the order of~$G$.
\end{lemma}

\begin{proof}
This is a straightforward corollary of Lemma \ref{lem:halving} excepting the case
that $d(u,v)=\infty$ while $d(u',v')$ is finite but strictly larger than $n$.
In this case in $G'$ there is a vertex $w'$ with $d(u',w')=n$.
Let Spoiler pebble $w'$. <Whatever Duplicator's response $w$ in $G$ is,
we have either $d(u,w)<n$ or $d(u,w)=\infty$ and Lemma \ref{lem:halving} applies
(with $G$ and $G'$ interchanged for the latter condition).
\end{proof}

We now show one of the directions in which the halving strategy can be developed.

\begin{lemma}{\bf(Generalized Halving Strategy)}\label{lem:exthalving}
\begin{bfenumerate}
\item
Let $z_1,\ldots,z_q,a_1,\ldots,a_m$ and $z'_1,\ldots,z'_q,a'_1,\ldots,a'_m$
be pairwise distinct vertices in graphs $G$ and $G'$ respectively. Suppose that,
for every $i<m$, Spoiler wins $\game^k_r(G,z_1,\ldots,z_q,a_i,a,G',z'_1,\ldots,z'_q,a'_i,a')$
whenever exactly one of the equalities $a=a_{i+1}$ and $a'=a'_{i+1}$ is true.
Then, for every $\ell\le m$, Spoiler wins 
$\game^K_R(G,z_1,\ldots,z_q,a_1,a_{\ell},G',z'_1,\ldots,z'_q,a'_1,a')$
whenever $a'\ne a'_{\ell}$, where $K=\max\{k,q+3\}$
and $R=\lceil\log\ell\rceil+r$.
\item
Let $t\ge1$ and each $a_i$ and $a'_i$ denote a tuple of $t$ vertices of $G$ and $G'$ respectively.
Then the preceding statement holds true with $K=\max\{k,q+3t\}$
and $R=\lceil\log\ell\rceil t+r$.
\end{bfenumerate}
\end{lemma}

\begin{proof}
We prove Item 1; Item 2 is a simple extension.
The overall idea can be explained as follows: We endow the vertex sets $V(G)$ and $V(G')$
with directed edges $(a_{i-1},a_i)$ and $(a'_{i-1},a'_i)$ for all $1<i\le m$ and
simulate the halving strategy on these directed graphs. However, compared to the proof
of Lemma \ref{lem:halving}, we have to ensure some special conditions for the endgame,
which requires some extra care.

Let $A=\{a_1,\ldots,a_m\}$ and $A'=\{a'_1,\ldots,a'_m\}$.
Suppose that $\ell>1$ for else the claim is trivial.

In our description of Spoiler's strategy in 
$\game^K_R(G,z_1,\ldots,z_q,a_1,a_{\ell},G',z'_1,\ldots,z'_q,a'_1,a')$,
we will denote the vertices of $G$ and $G'$ pebbled in the $s$-th round
by $w_s$ and $w'_s$ respectively. Simultaneously we will define auxiliary vertices
$u_s,v_s\in V(G)$ and $u'_s,v'_s\in V(G')$. Spoiler's choice of $w_s$ or $w'_s$
will depend on the vertices $u_{s-1}$, $v_{s-1}$, $u'_{s-1}$, and $v'_{s-1}$, which all will be
pebbled by the beginning of the $s$-th round.
Initially we set $u_0=a_1$, $v_0=a_\ell$, $u'_0=a'_1$, and $v'_0=a'$.
We will take care that, for every $s$, the following is true.
Tere are three integers $1\le i_s<j_s<h_s\le m$ such that
$u_s=a_{i_s}$, $u'_s=a'_{i_s}$, and the pair $v_s,v'_s$ satisfies one of the
following two conditions:
$$
v_s=a_{j_s}\mbox{\ and\ either\ }v'_s=a'_{h_s}\mbox{\ or\ }v'_s\notin A'
\leqno (C)
$$
or
$$
v'_s=a'_{j_s}\mbox{\ and\ either\ }v_s=a_{h_s}\mbox{\ or\ }v_s\notin A.
\leqno (C')
$$
Note that for $s=0$ these properties are true.

Suppose that $u_{s-1}$, $v_{s-1}$, $u'_{s-1}$, and $v'_{s-1}$ are defined
and have the avbove properties. This specifies the integers $i_{s-1}$, $j_{s-1}$,
and $h_{s-1}$ (the latter may be undefined). Assume that $j_{s-1}>i_{s-1}+1$.
We describe Spoiler's move and the new quadruple $(u_s,v_s,u'_s,v'_s)$ under
Condition $(C)$. For Condition $(C')$ the description is symmetric.
Let $t=\lfloor(i_{s-1}+j_{s-1})/2\rfloor$. Spoiler pebbles $w_s=a_t$.

If for Duplicator's response $w'_s$ we have $w'_s\notin A'$, then set
\begin{equation}\label{eq:uvs}
u_s=u_{s-1},\ u'_s=u'_{s-1},\ v_s=w_s,\ v'_s=w'_s.
\end{equation}
Note that the quadruple $(u_s,v_s,u'_s,v'_s)$ has the required properties,
in particular, Condition $(C)$ holds true.
The pebbles on $v_{s-1}$ and $v'_{s-1}$ can be now released and reused.

Suppose now that $w'_s=a'_p$ for some $p\le m$. If $p=t$, then set
\begin{equation}
u_s=w_s,\ u'_s=w'_s,\ v_s=v_{s-1},\ v'_s=v'_{s-1}.
\end{equation}
Note that $(C)$ holds true. If $p\ne t$, apply the setting \refeq{eq:uvs}.
If $p>t$, then $(C)$ holds true; if $p<t$, then $(C')$ is true.

Notice that $j_s-i_s\le\lceil(j_{s-1}-i_{s-1})/2\rceil$.
It follows that at latest for $s=\lceil\log\ell\rceil$ we get $j_s=i_s+1$.
Spoiler needs no more than $r$ extra moves to win starting from this point
because, by the assumption, Spoiler has a winning strategy in
$\game^k_r(G,z_1,\ldots,z_q,u_s,v_s,G',z'_1,\ldots,z'_q,u'_s,v'_s)$.
\end{proof}

\section{Local strategy}\label{s:local}

We start with proving Theorem \ref{thm:dist3}.
We have to design a strategy allowing Spoiler to win the \EF\/ game on $G$ and $G'$
with 15 pebbles in $O(\log n)$ rounds. A crucial fact on which the strategy will be
based is the rigidity of triconnected planar graphs as stated in the Whitney theorem.
In this section we aim at developing an important ingredient of the strategy
forcing Duplicator to respect this rigidity. 

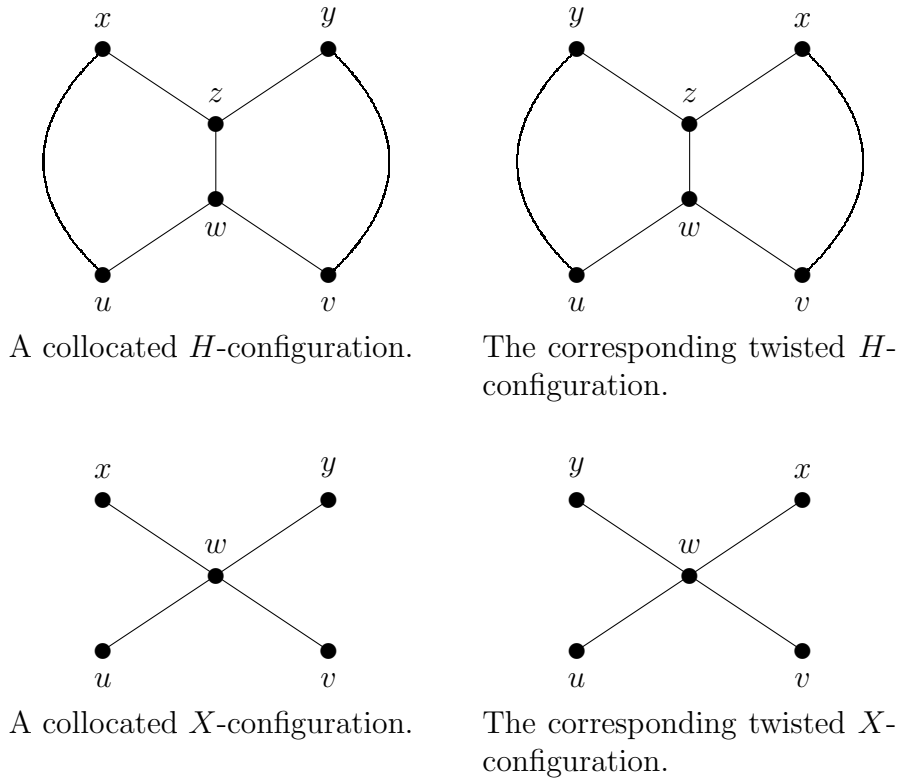
\begin{figure}[htbp]
\centerline{
\unitlength=1.00mm
\special{em:linewidth 0.4pt}
\linethickness{0.4pt}
\begin{picture}(134.00,98.00)
\put(25.00,95.00){\circle*{2.00}}
\put(40.00,85.00){\circle*{2.00}}
\put(55.00,95.00){\circle*{2.00}}
\put(40.00,75.00){\circle*{2.00}}
\put(25.00,65.00){\circle*{2.00}}
\put(55.00,65.00){\circle*{2.00}}
\emline{25.00}{95.00}{1}{40.00}{85.00}{2}
\emline{40.00}{85.00}{3}{40.00}{75.00}{4}
\emline{40.00}{75.00}{5}{25.00}{65.00}{6}
\emline{40.00}{85.00}{7}{55.00}{95.00}{8}
\emline{40.00}{75.00}{9}{55.00}{65.00}{10}
\bezier{176}(25.00,95.00)(9.00,80.00)(25.00,65.00)
\bezier{176}(55.00,95.00)(71.00,80.00)(55.00,65.00)
\put(25.00,98.00){\makebox(0,0)[cb]{$x$}}
\put(55.00,98.00){\makebox(0,0)[cb]{$y$}}
\put(55.00,62.00){\makebox(0,0)[ct]{$v$}}
\put(25.00,62.00){\makebox(0,0)[ct]{$u$}}
\put(40.00,88.00){\makebox(0,0)[cb]{$z$}}
\put(40.00,72.00){\makebox(0,0)[ct]{$w$}}
\put(40.00,57.00){\makebox(0,0)[ct]{\parbox{55mm}{A collocated $H$-configuration.}}}
\put(88.00,95.00){\circle*{2.00}}
\put(103.00,85.00){\circle*{2.00}}
\put(118.00,95.00){\circle*{2.00}}
\put(103.00,75.00){\circle*{2.00}}
\put(88.00,65.00){\circle*{2.00}}
\put(118.00,65.00){\circle*{2.00}}
\emline{88.00}{95.00}{11}{103.00}{85.00}{12}
\emline{103.00}{85.00}{13}{103.00}{75.00}{14}
\emline{103.00}{75.00}{15}{88.00}{65.00}{16}
\emline{103.00}{85.00}{17}{118.00}{95.00}{18}
\emline{103.00}{75.00}{19}{118.00}{65.00}{20}
\bezier{176}(88.00,95.00)(72.00,80.00)(88.00,65.00)
\bezier{176}(118.00,95.00)(134.00,80.00)(118.00,65.00)
\put(88.00,98.00){\makebox(0,0)[cb]{$y$}}
\put(118.00,98.00){\makebox(0,0)[cb]{$x$}}
\put(118.00,62.00){\makebox(0,0)[ct]{$v$}}
\put(88.00,62.00){\makebox(0,0)[ct]{$u$}}
\put(103.00,88.00){\makebox(0,0)[cb]{$z$}}
\put(103.00,72.00){\makebox(0,0)[ct]{$w$}}
\put(103.00,57.00){\makebox(0,0)[ct]{\parbox{55mm}{The corresponding twisted $H$-configuration.}}}
\put(25.00,35.00){\circle*{2.00}}
\put(40.00,25.00){\circle*{2.00}}
\put(55.00,35.00){\circle*{2.00}}
\emline{25.00}{35.00}{21}{40.00}{25.00}{22}
\emline{40.00}{25.00}{23}{55.00}{35.00}{24}
\put(25.00,38.00){\makebox(0,0)[cb]{$x$}}
\put(55.00,38.00){\makebox(0,0)[cb]{$y$}}
\put(40.00,28.00){\makebox(0,0)[cb]{$w$}}
\put(25.00,15.00){\circle*{2.00}}
\put(55.00,15.00){\circle*{2.00}}
\emline{40.00}{25.00}{25}{25.00}{15.00}{26}
\emline{40.00}{25.00}{27}{55.00}{15.00}{28}
\put(55.00,12.00){\makebox(0,0)[ct]{$v$}}
\put(25.00,12.00){\makebox(0,0)[ct]{$u$}}
\put(40.00,7.00){\makebox(0,0)[ct]{\parbox{55mm}{A collocated $X$-configuration.}}}
\put(88.00,35.00){\circle*{2.00}}
\put(103.00,25.00){\circle*{2.00}}
\put(118.00,35.00){\circle*{2.00}}
\emline{88.00}{35.00}{29}{103.00}{25.00}{30}
\emline{103.00}{25.00}{31}{118.00}{35.00}{32}
\put(88.00,38.00){\makebox(0,0)[cb]{$y$}}
\put(118.00,38.00){\makebox(0,0)[cb]{$x$}}
\put(103.00,28.00){\makebox(0,0)[cb]{$w$}}
\put(88.00,15.00){\circle*{2.00}}
\put(118.00,15.00){\circle*{2.00}}
\emline{103.00}{25.00}{33}{88.00}{15.00}{34}
\emline{103.00}{25.00}{35}{118.00}{15.00}{36}
\put(118.00,12.00){\makebox(0,0)[ct]{$v$}}
\put(88.00,12.00){\makebox(0,0)[ct]{$u$}}
\put(103.00,7.00){\makebox(0,0)[ct]{\parbox{55mm}{The corresponding twisted $X$-configuration.}}}
\end{picture}
}
\caption{Definition of a configuration}%
\label{fig:1}
\end{figure}

A \emph{configuration} $C$ in a graph $G$ is a set of labeled vertices of $G$.
In fact, labels will be the pebbles in an \EF\/ game. 
At the same time we will often use a label as a name of a vertex.
By an \emph{$X$-configuration} we mean 5 pairwise distinct vertices labeled by $x,y,u,v$, and $w$
such that $x,y,u,v\in\Gamma(w)$.
By an \emph{$H$-configuration} we mean 6 pairwise distinct vertices labeled by $x,y,z,u,v$, and $w$
such that $z$ and $w$ are adjacent, $x,y\in\Gamma(z)$, and $u,v\in\Gamma(w)$.
Thus, contraction of the edge $\{z,w\}$ makes an $H$-configuration an $X$-configuration.
Suppose that $G$ is a triconnected planar graph and consider its unique spherical
embedding. We call an $X$-configuration $C$ \emph{collocated} if  $u,x,y,v$ occur around $w$
exactly in this order (up to cyclic shifts and the direction of a roundabout way).
We call an $H$-configuration $C$ \emph{collocated} if $xzwu$ and $yzwv$ are seqments
of the two facial cycles containing the edge $\{z,w\}$.
A configuration obtained from a collocated $X$- or $H$-configuration by
interchanging the labels $x$ and $y$
will be called a \emph{twisted} configuration, see Figure~\ref{fig:1}.

We will treat $X$- and $H$-configurations uniformly, setting $z=w$ for $X$-configurations.
Whenever we use the term \emph{configuration} alone, it will refer to
any $X$- or $H$-configuration.

\begin{lemma}\label{lem:colltwist}
Let $G$ and $G'$ be triconnected planar graphs, $G$ having $n$ vertices.
Let $C=\{x,y,z,u,v,w\}$ and $C'=\{x',y',z',u',v',w'\}$ be sets of pebbled vertices
in, respectively, $G$ and $G'$ such that $C$ is a collocated configuration and
$C'$ is a twisted configuration. Starting with this position, Spoiler wins
the \EF\/ game on $G$ and $G'$ with 
15 pebbles in less than $6\log n+26$ moves.
\end{lemma}

The proof occupies the rest of the section.

\subsection{Preliminaries I}

\begin{definition}\label{def:distavoid}\rm
Suppose that in a graph $G$ a configuration $C$ is designated.
We write $d_0(a,b)$ to denote the minimum
length of an $a$-$b$-path avoiding $C$. Note that $a$ and $b$
may be arbitrary, in particular, belong to $C$.
Furthermore, $\calS_0(a,b)$ will denote the set of all $a$-$b$-paths that 
avoid $C$ and have length $d_0(a,b)$, that is, are as short as possible.
A path in $\calS_0(a,b)$ will be called a \emph{$d_0$-geodesics} from $a$ to $b$.
Finally, $S_0(a,b)$ will denote the set of all vertices belonging to at least
one $d_0$-geodesics from $a$ to~$b$.
\end{definition}

Note that $d_0$ does not necessary satisfy the triangle inequality, though
most basic properties $d_0(a,a)=0$ and $d_0(a,b)=d_0(b,a)$ are preserved.

\begin{lemma}\label{lem:Snonempty}
If $C$ is a collocated configuration in a triconnected planar graph $G$,
then $\calS_0(x,u)\ne\emptyset$, that is, $d_0(x,u)<\infty$.
The same holds true for $y$ and $v$.
\footnote{This is actually a logical consequence of the claim about $x$ and~$u$.}
\end{lemma}

\begin{figure}[htbp]
\centerline{
\unitlength=1.00mm
\special{em:linewidth 0.4pt}
\linethickness{0.4pt}
\begin{picture}(153.00,79.00)
\put(25.00,40.00){\circle*{2.00}}
\put(40.00,30.00){\circle*{2.00}}
\put(55.00,40.00){\circle*{2.00}}
\put(40.00,20.00){\circle*{2.00}}
\put(25.00,10.00){\circle*{2.00}}
\put(55.00,10.00){\circle*{2.00}}
\emline{25.00}{40.00}{1}{40.00}{30.00}{2}
\emline{40.00}{30.00}{3}{40.00}{20.00}{4}
\emline{40.00}{20.00}{5}{25.00}{10.00}{6}
\emline{40.00}{30.00}{7}{55.00}{40.00}{8}
\emline{40.00}{20.00}{9}{55.00}{10.00}{10}
\put(25.00,43.00){\makebox(0,0)[cb]{$x$}}
\put(55.00,43.00){\makebox(0,0)[cb]{$y$}}
\put(55.00,7.00){\makebox(0,0)[ct]{$v$}}
\put(25.00,7.00){\makebox(0,0)[ct]{$u$}}
\put(40.00,33.00){\makebox(0,0)[cb]{$z$}}
\put(40.00,17.00){\makebox(0,0)[ct]{$w$}}
\bezier{540}(55.00,40.00)(5.00,79.00)(25.00,10.00)
\bezier{124}(25.00,40.00)(43.00,36.00)(55.00,40.00)
\put(105.00,40.00){\circle*{2.00}}
\put(135.00,40.00){\circle*{2.00}}
\put(105.00,10.00){\circle*{2.00}}
\put(135.00,10.00){\circle*{2.00}}
\put(105.00,43.00){\makebox(0,0)[cb]{$x$}}
\put(135.00,43.00){\makebox(0,0)[cb]{$y$}}
\put(135.00,7.00){\makebox(0,0)[ct]{$v$}}
\put(105.00,7.00){\makebox(0,0)[ct]{$u$}}
\put(120.00,25.00){\circle*{2.00}}
\emline{105.00}{10.00}{11}{135.00}{40.00}{12}
\emline{105.00}{40.00}{13}{135.00}{10.00}{14}
\put(123.00,25.00){\makebox(0,0)[lc]{$w$}}
\put(105.00,30.00){\circle*{2.00}}
\put(105.00,20.00){\circle*{2.00}}
\emline{105.00}{20.00}{15}{120.00}{25.00}{16}
\emline{120.00}{25.00}{17}{105.00}{30.00}{18}
\bezier{368}(105.00,30.00)(86.00,63.00)(135.00,40.00)
\bezier{184}(135.00,40.00)(153.00,35.00)(135.00,55.00)
\bezier{588}(135.00,55.00)(64.00,79.00)(105.00,20.00)
\end{picture}
}
\caption{Proof of Lemma \protect\ref{lem:Snonempty}}%
\label{fig:2}
\end{figure}

\begin{proof}
Suppose that $C$ is an $H$-configuration.
Let $F$ be the facial cycle going through $x,z,w,u$.
Note that $F$ does not go through $y$ (nor, similarly, through $v$)
because otherwise $G\setminus\{z,y\}$ would be disconnected, see Figure \ref{fig:2}.
Thus, removal of $z$ and $w$ from $F$ gives us an $x$-$u$-path avoiding $C$ 
and hence $\calS_0(x,u)\ne\emptyset$.

Suppose now that $C$ is an $X$-configuration.
The case that $x,w,u$ lie on a facial cycle is completely similar to the case 
of an $H$-configuration. In the general case our argument is a bit longer.
We say that a facial cycle $F$ going through $w$ is \emph{between} $x$ and $u$
if both neighbors of $w$ in $F$ lie between $x$ and $u$ in the embedding of $G$.
To make the notion of \emph{betweenness} unambiguous, we shall agree that
$y$ and $v$ are \emph{not} between $x$ and $u$. Let $F$ be a facial cycle between $x$ and $u$.
Again, $F$ does not go through $y$ (nor, similarly, through $v$)
because otherwise $G\setminus\{z,y\}$ would be disconnected, see Figure \ref{fig:2}.
Consider the sum of all facial cycles between $x$ and $u$ (it consists of the edges
occurring in only one of the summands; if an edge appears in two summands, it is eliminated).
This is a cycle and removal of $w$ from it gives us a path in $\calS_0(x,u)$.
\end{proof}

\begin{definition}\rm
Let $\calP$ and $\calQ$ be sets of paths. We say that $\calP$ and $\calQ$ have 
\begin{itemize}
\item
\emph{intersection property} if for every $P\in\calP$ there is $Q\in\calQ$
and for every $Q\in\calQ$ there is $P\in\calP$ such that $P$ and $Q$
have a common inner point;
\item
\emph{strong intersection property} if every $P\in\calP$ and $Q\in\calQ$
have a common inner point.
\end{itemize}
\end{definition}

\begin{lemma}\label{lem:twstrong}
If $C$ is a twisted $X$- or $H$-configuration and both $\calS_0(x,u)$ and $\calS_0(y,v)$
are non-empty, then these path sets
have the strong intersection property.\noproof
\end{lemma}

\subsection{The case of $\calS_0(x,u)$ and $\calS_0(y,v)$ not having the
intersection property}\label{ss:notint}

The proof of Lemma \ref{lem:colltwist} is based on Proposition \ref{prop:game}.
We have to analyse the \EF\/ game on non-isomorphic triconnected planar graphs $G$ and $G'$
starting from a position where configurations $C$ and $C'$ are pebbled in,
respectively, $G$ and $G'$ so that $C$ is collocated while $C'$ is twisted.
The latter precondition is default in the sequel.

\begin{lemma}\label{lem:d0respect}
Duplicator is forced to respect the distance $d_0$: 
Once vertices $a,b\in V(G)$ and $a',b'\in V(G')$
are pebbled so that $d_0(a,b)\ne d_0(a',b')$,
Spoiler wins with 9 pebbles in less than $\log n+2$ moves.
\end{lemma}

\begin{proof}
The inequality $d_0(a,b)\ne d_0(a',b')$ means that
$d(a,b)\ne d(a',b')$ in graphs $G\setminus(C\setminus\{a,b\})$
and $G'\setminus(C'\setminus\{a',b'\})$. We are hence done by Lemma \ref{lem:drespect}. 
Note that Spoiler operates with 3 new pebbles while 6 pebbles must all the time
remain on~$C$.
\end{proof}

Lemma \ref{lem:d0respect} applies to the case that
$d_0(x',u')\ne d_0(x,u)$ or $d_0(y',v')\ne d_0(y,v)$.
Whenever we show that Spoiler has an efficient winning strategy under a certain
condition, for the rest we will make an assumption that this condition is not met.

\begin{assumption}\label{ass:eqdist}\rm
We have $d_0(x',u')=d_0(x,u)$ and $d_0(y',v')=d_0(y,v)$.
In particular, both $\calS_0(x',u')$ and $\calS_0(y',v')$ are non-empty
(as a consequence of Lemma~\ref{lem:Snonempty}).
\end{assumption}

\begin{lemma}\label{lem:srespect}
Let $s,t\in V(G)$ and $s',t'\in V(G')$ be pebbled so that $d_0(s,t)=d_0(s',t')$.
Then Duplicator is forced to respect $S_0(s,t)$ and $S_0(s',t')$: 
Once $a\in V(G)$ and $a'\in V(G')$ are pebbled so that $a\in S_0(s,t)$ but
$a'\notin S_0(s',t')$ or vice versa,
Spoiler wins with 9 pebbles in $\log n+2$ moves.
\end{lemma}

\begin{proof}
Suppose that $a\in C$ or $a'\in C'$; otherwise the claim trivializes.
Note that $a\in S_0(s,t)$ iff $d_0(s,a)+d_0(a,t)=d_0(s,t)$;
and $S_0(s',t')$ has a similar characterization. By assumptions, we therefore
have either
$d_0(s,a)\ne d_0(s',a')$ or $d_0(a,t)\ne d_0(a',t')$ and Lemma \ref{lem:d0respect} applies.
\end{proof}

Below, the statement of each lemma begins with explicitly listing assumptions
used in its proof.

\begin{lemma}{\assfont[Assumption \ref{ass:eqdist}]}
Suppose that $\calS_0(x,u)$ and $\calS_0(y,v)$ do not have the intersection
property. Then Spoiler wins with 9 pebbles in $2\log n+2$ moves.
\end{lemma}

\begin{proof}
Assume, for example, that there is a $P\in\calS_0(x,u)$ avoiding $S_0(y,v)$.
By Lemma \ref{lem:twstrong}, no path in $\calS_0(x',u')$ can avoid $S_0(y',v')$.
This means that in graphs $H=G\setminus(\{z,w\}\cup S_0(y,v))$ and 
$H'=G'\setminus(\{z',w'\}\cup S_0(y',v'))$ we have $d(x,u)=d_0(x,u)$ while $d(x',u')>d_0(x',u')$,
and hence $d(x,u)<d(x',u')$.
and Spoiler employs the halving strategy of Lemma \ref{lem:halving} for these graphs.
To force play on $H$ and $H'$, Spoiler never moves in $\{z,w\}\cup S_0(y,v)$.
Once Duplicator deviates from playing in $H'$ and makes a move in $S_0(y',v')$,
she loses by Lemma \ref{lem:srespect}.
\end{proof}

\begin{assumption}\label{ass:intersect}\rm
$\calS_0(x,u)$ and $\calS_0(y,v)$ have the intersection property.
\end{assumption}

\subsection{Preliminaries II}

All considerations below for the pair $xu$ carry over to $yv$ by symmetry.

\begin{definition}\label{def:boundary}\rm
Let $C$ be an $X$- or $H$-configuration in a sphere graph $G$.
Given $P\in\calS_0(x,u)$, let $\bar P$ denote the cycle $uPxzw$.
We call $P$ a \emph{boundary} of $\calS_0(x,u)$ if all $S_0(x,u)$ lies within the same region
(one of the two) bounded by~$\bar P$.
\end{definition}

\begin{lemma}\label{lem:twobounds}
Let $C$ be an $X$- or $H$-configuration in a sphere graph $G$.
If $|\calS_0(x,u)|>1$, then $\calS_0(x,u)$ has exactly two boundary paths.
\end{lemma}

\begin{figure}[htbp]
\centerline{
\unitlength=1.00mm
\special{em:linewidth 0.4pt}
\linethickness{0.4pt}
\begin{picture}(131.00,43.00)
\put(25.00,40.00){\circle*{2.00}}
\put(40.00,30.00){\circle*{2.00}}
\put(55.00,40.00){\circle*{2.00}}
\put(40.00,20.00){\circle*{2.00}}
\put(25.00,10.00){\circle*{2.00}}
\put(55.00,10.00){\circle*{2.00}}
\emline{25.00}{40.00}{1}{40.00}{30.00}{2}
\emline{40.00}{30.00}{3}{40.00}{20.00}{4}
\emline{40.00}{20.00}{5}{25.00}{10.00}{6}
\emline{40.00}{30.00}{7}{55.00}{40.00}{8}
\emline{40.00}{20.00}{9}{55.00}{10.00}{10}
\put(25.00,43.00){\makebox(0,0)[cb]{$x$}}
\put(55.00,43.00){\makebox(0,0)[cb]{$y$}}
\put(55.00,7.00){\makebox(0,0)[ct]{$v$}}
\put(25.00,7.00){\makebox(0,0)[ct]{$u$}}
\put(40.00,33.00){\makebox(0,0)[cb]{$z$}}
\put(40.00,17.00){\makebox(0,0)[ct]{$w$}}
\bezier{124}(25.00,40.00)(22.00,25.00)(25.00,10.00)
\bezier{56}(25.00,40.00)(18.00,39.00)(24.00,35.00)
\bezier{112}(24.00,35.00)(36.00,29.00)(24.00,21.00)
\bezier{96}(24.00,21.00)(14.00,15.00)(25.00,10.00)
\put(24.00,35.00){\circle*{2.00}}
\put(24.00,21.00){\circle*{2.00}}
\put(30.00,27.00){\circle*{2.00}}
\put(21.00,34.00){\makebox(0,0)[rt]{$a$}}
\put(33.00,27.00){\makebox(0,0)[lc]{$s$}}
\put(21.00,21.00){\makebox(0,0)[rb]{$b$}}
\put(100.00,40.00){\circle*{2.00}}
\put(115.00,30.00){\circle*{2.00}}
\put(130.00,40.00){\circle*{2.00}}
\put(115.00,20.00){\circle*{2.00}}
\put(100.00,10.00){\circle*{2.00}}
\put(130.00,10.00){\circle*{2.00}}
\emline{100.00}{40.00}{11}{115.00}{30.00}{12}
\emline{115.00}{30.00}{13}{115.00}{20.00}{14}
\emline{115.00}{20.00}{15}{100.00}{10.00}{16}
\emline{115.00}{30.00}{17}{130.00}{40.00}{18}
\emline{115.00}{20.00}{19}{130.00}{10.00}{20}
\put(100.00,43.00){\makebox(0,0)[cb]{$x$}}
\put(130.00,43.00){\makebox(0,0)[cb]{$y$}}
\put(130.00,7.00){\makebox(0,0)[ct]{$v$}}
\put(100.00,7.00){\makebox(0,0)[ct]{$u$}}
\put(115.00,33.00){\makebox(0,0)[cb]{$z$}}
\put(115.00,17.00){\makebox(0,0)[ct]{$w$}}
\bezier{108}(100.00,40.00)(89.00,32.00)(100.00,25.00)
\bezier{100}(100.00,40.00)(110.00,32.00)(100.00,25.00)
\bezier{108}(100.00,25.00)(89.00,17.00)(100.00,10.00)
\bezier{100}(100.00,25.00)(110.00,17.00)(100.00,10.00)
\bezier{64}(100.00,40.00)(98.00,32.00)(100.00,25.00)
\bezier{64}(100.00,25.00)(103.00,16.00)(100.00,10.00)
\end{picture}
}
\caption{Proof of Lemma \protect\ref{lem:twobounds}}%
\label{fig:3}
\end{figure}

\begin{proof}
Call $P\in\calS_0(x,u)$ a \emph{border} of $\calS_0(x,u)$ if the bounded region of $\bar P$
contains the smallest possible number of points from $S_0(x,u)$.
The latter number is actually equal to 0. Indeed, suppose that
$P$ is a border and the bounded region of $\bar P$ contains a point $s\in S_0(x,u)$.
Let $Q$ be a path in $\calS_0(x,u)$ going through $s$. Denote the two common vertices of 
$Q$ and $P$ which are nearest to $s$ in $Q$ by $a$ and $b$. Replace the segment $P[a,b]$
of $P$ by $Q[a,b]$. Since both $P[a,b]$ and $Q[a,b]$ have length $d_0(a,b)$,
the modified path $R$ is in $\calS_0(x,u)$. The bounded region of $\bar R$
does not contain $s$ and cannot have any new point because it is
included in the bounded region of $\bar P$, a contradiction (see Figure \ref{fig:3}).

As easily follows from the definitions, a border of $\calS_0(x,u)$ is a boundary of $\calS_0(x,u)$.
As a matter of definition, a border exists and we hence have proved
the existence of one boundary. The existence of the second boundary follows 
from the same argument if instead of bounded regions we consider unbounded regions.
Since for any third path $Q\in\calS_0(x,u)$, the cycle $\bar Q$ cannot surround
these two boundary paths simultaneously, no third boundary exists, see Figure~\ref{fig:3}.
\end{proof}

\begin{convention}\rm
We will apply Definition \ref{def:boundary} to a triconnected planar graph $G$
assuming its (unique) sphere embedding. 
It is not hard to deduce from the Whitney theorem that the
boundary paths do not depend a particular embedding.
\end{convention}

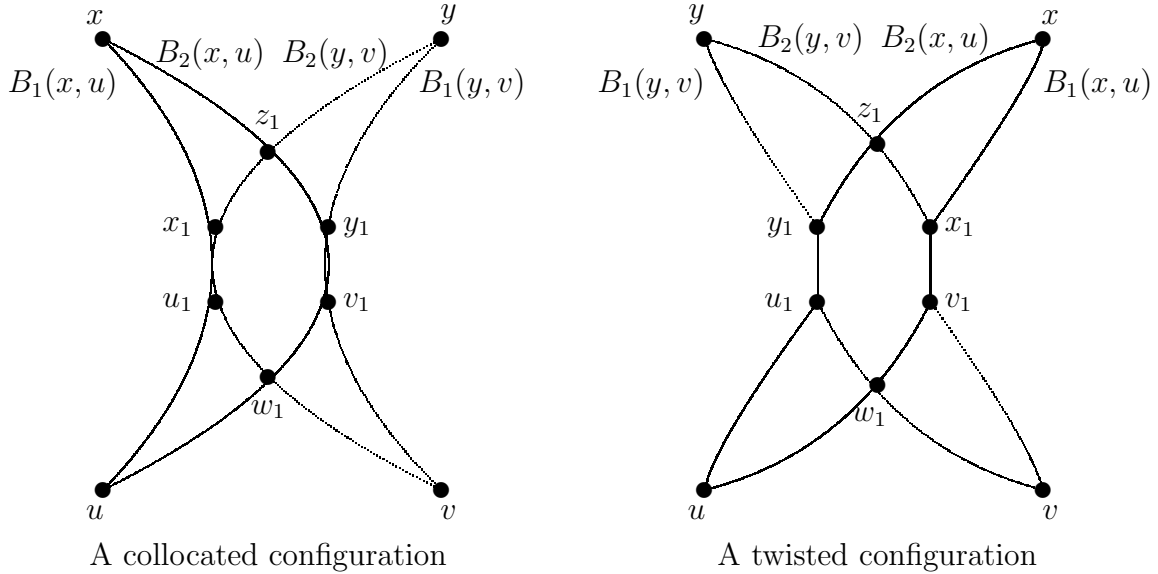
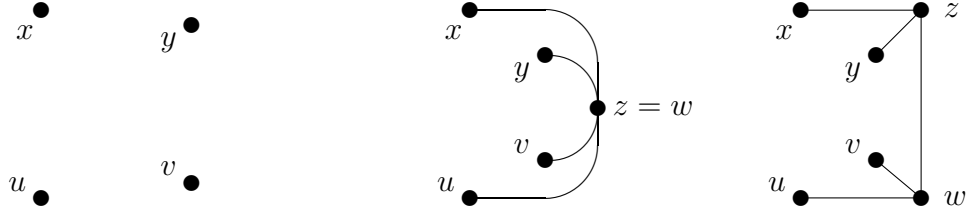
\begin{figure}[htbp]
\centerline{
\unitlength=1.00mm
\special{em:linewidth 0.4pt}
\linethickness{0.4pt}
\begin{picture}(136.00,138.00)
\put(82.50,124.00){\oval(13.00,14.00)[r]}
\put(72.50,124.50){\oval(33.00,25.00)[r]}
\put(72.00,137.00){\circle*{2.00}}
\put(72.00,112.00){\circle*{2.00}}
\put(82.00,131.00){\circle*{2.00}}
\put(82.00,117.00){\circle*{2.00}}
\put(89.00,124.00){\circle*{2.00}}
\put(71.00,135.00){\makebox(0,0)[rt]{$x$}}
\put(70.00,113.00){\makebox(0,0)[rb]{$u$}}
\put(80.00,130.00){\makebox(0,0)[rt]{$y$}}
\put(80.00,118.00){\makebox(0,0)[rb]{$v$}}
\put(91.00,124.00){\makebox(0,0)[lc]{$z=w$}}
\put(116.00,137.00){\circle*{2.00}}
\put(116.00,112.00){\circle*{2.00}}
\put(126.00,131.00){\circle*{2.00}}
\put(126.00,117.00){\circle*{2.00}}
\put(115.00,135.00){\makebox(0,0)[rt]{$x$}}
\put(114.00,113.00){\makebox(0,0)[rb]{$u$}}
\put(124.00,130.00){\makebox(0,0)[rt]{$y$}}
\put(124.00,118.00){\makebox(0,0)[rb]{$v$}}
\put(132.00,137.00){\circle*{2.00}}
\put(132.00,112.00){\circle*{2.00}}
\emline{116.00}{112.00}{1}{132.00}{112.00}{2}
\emline{132.00}{112.00}{3}{132.00}{137.00}{4}
\emline{132.00}{137.00}{5}{116.00}{137.00}{6}
\emline{132.00}{137.00}{7}{126.00}{131.00}{8}
\emline{132.00}{112.00}{9}{126.00}{117.00}{10}
\put(135.00,137.00){\makebox(0,0)[lc]{$z$}}
\put(135.00,112.00){\makebox(0,0)[lc]{$w$}}
\put(15.00,137.00){\circle*{2.00}}
\put(15.00,112.00){\circle*{2.00}}
\put(14.00,135.00){\makebox(0,0)[rt]{$x$}}
\put(13.00,113.00){\makebox(0,0)[rb]{$u$}}
\put(35.00,135.00){\circle*{2.00}}
\put(33.00,134.00){\makebox(0,0)[rt]{$y$}}
\put(35.00,114.00){\circle*{2.00}}
\put(33.00,115.00){\makebox(0,0)[rb]{$v$}}
\put(5.00,104.00){\makebox(0,0)[lt]{\parbox{130mm}{Depicting an $X$- and an $H$-configuration in a uniform way. The leftmost picture can be a fragment of either of the two others.}}}
\put(10.00,77.00){\circle*{2.00}}
\put(55.00,77.00){\circle*{2.00}}
\put(55.00,17.00){\circle*{2.00}}
\put(10.00,17.00){\circle*{2.00}}
\bezier{200}(55.00,77.00)(24.00,47.00)(55.00,17.00)
\bezier{650}(10.00,77.00)(70.00,47.00)(10.00,17.00)
\put(25.00,52.00){\circle*{2.00}}
\put(25.00,42.00){\circle*{2.00}}
\put(40.00,42.00){\circle*{2.00}}
\put(40.00,52.00){\circle*{2.00}}
\put(32.00,62.00){\circle*{2.00}}
\put(32.00,32.00){\circle*{2.00}}
\bezier{250}(55.00,17.00)(-6.00,47.00)(55.00,77.00)
\bezier{600}(10.00,17.00)(39.00,47.00)(10.00,77.00)
\put(9.00,79.00){\makebox(0,0)[cb]{$x$}}
\put(56.00,79.00){\makebox(0,0)[cb]{$y$}}
\put(56.00,15.00){\makebox(0,0)[ct]{$v$}}
\put(9.00,15.00){\makebox(0,0)[ct]{$u$}}
\put(22.00,52.00){\makebox(0,0)[rc]{$x_1$}}
\put(22.00,42.00){\makebox(0,0)[rc]{$u_1$}}
\put(42.00,42.00){\makebox(0,0)[lc]{$v_1$}}
\put(42.00,52.00){\makebox(0,0)[lc]{$y_1$}}
\put(32.00,65.00){\makebox(0,0)[cb]{$z_1$}}
\put(32.00,29.00){\makebox(0,0)[ct]{$w_1$}}
\put(12.00,71.00){\makebox(0,0)[rc]{$B_1(x,u)$}}
\put(17.00,75.00){\makebox(0,0)[lc]{$B_2(x,u)$}}
\put(48.00,75.00){\makebox(0,0)[rc]{$B_2(y,v)$}}
\put(52.00,71.00){\makebox(0,0)[lc]{$B_1(y,v)$}}
\put(90.00,77.00){\circle*{2.00}}
\put(135.00,77.00){\circle*{2.00}}
\put(135.00,17.00){\circle*{2.00}}
\put(90.00,17.00){\circle*{2.00}}
\put(105.00,52.00){\circle*{2.00}}
\put(105.00,42.00){\circle*{2.00}}
\put(120.00,42.00){\circle*{2.00}}
\put(120.00,52.00){\circle*{2.00}}
\put(89.00,79.00){\makebox(0,0)[cb]{$y$}}
\put(136.00,79.00){\makebox(0,0)[cb]{$x$}}
\put(136.00,15.00){\makebox(0,0)[ct]{$v$}}
\put(89.00,15.00){\makebox(0,0)[ct]{$u$}}
\put(102.00,52.00){\makebox(0,0)[rc]{$y_1$}}
\put(102.00,42.00){\makebox(0,0)[rc]{$u_1$}}
\put(122.00,42.00){\makebox(0,0)[lc]{$v_1$}}
\put(122.00,52.00){\makebox(0,0)[lc]{$x_1$}}
\bezier{100}(105.00,52.00)(105.00,47.00)(105.00,42.00)
\bezier{100}(120.00,52.00)(120.00,47.00)(120.00,42.00)
\bezier{320}(120.00,52.00)(135.00,73.00)(135.00,77.00)
\bezier{130}(120.00,52.00)(110.00,72.00)(90.00,77.00)
\bezier{420}(135.00,77.00)(115.00,72.00)(105.00,52.00)
\bezier{420}(90.00,17.00)(110.00,22.00)(120.00,42.00)
\bezier{130}(135.00,17.00)(115.00,22.00)(105.00,42.00)
\bezier{80}(105.00,52.00)(92.00,70.00)(90.00,77.00)
\bezier{320}(105.00,42.00)(91.00,23.00)(90.00,17.00)
\bezier{80}(135.00,17.00)(133.00,24.00)(120.00,42.00)
\put(113.00,63.00){\circle*{2.00}}
\put(113.00,31.00){\circle*{2.00}}
\put(112.00,66.00){\makebox(0,0)[cb]{$z_1$}}
\put(112.00,28.00){\makebox(0,0)[ct]{$w_1$}}
\put(90.00,71.00){\makebox(0,0)[rc]{$B_1(y,v)$}}
\put(135.00,71.00){\makebox(0,0)[lc]{$B_1(x,u)$}}
\put(97.00,77.00){\makebox(0,0)[lc]{$B_2(y,v)$}}
\put(128.00,77.00){\makebox(0,0)[rc]{$B_2(x,u)$}}
\put(32.00,10.00){\makebox(0,0)[ct]{A collocated configuration}}
\put(113.00,10.00){\makebox(0,0)[ct]{A twisted configuration}}
\end{picture}
}
\caption{The $d_0$-roadmap. Warning: $B_1(x,u)$ and $B_2(x,u)$ (as well as $B_1(y,v)$ and $B_2(y,v)$)
can actually intersect a number of times.}%
\label{fig:4}
\end{figure}

\begin{convention}\rm
If $|\calS_0(x,u)|\ge2$, then the two boundaries of $\calS_0(x,u)$ will be denoted
by $B_1(x,u)$ and $B_2(x,u)$. Furthermore, we fix the indices so that
$B_2(x,u)$ is in the same region of $\bar B_1(x,u)$ together with $y$.
The similar notation will be used for the pair $y,v$. In this case
the indices are fixed so that
$B_2(y,v)$ is in the same region of $\bar B_1(y,v)$ together with $x$, see Figure~\ref{fig:4}.

If $|\calS_0(x,u)|=1$, we let $B_1(x,u)=B_2(x,u)$ denote
the single path in $\calS_0(x,u)$. 
\end{convention}

Assumption \ref{ass:intersect} can now be rephrased so that 
$B_1(x,u)$ and $B_2(y,v)$ as well as $B_1(y,v)$ and $B_2(x,u)$ touch at some points.

\begin{definition}\label{def:C1}\rm
We fix notation as in~Figure~\ref{fig:4}.

$z_1$ denotes the common vertex of $B_2(x,u)$ and $B_2(y,v)$ nearest to $x$ in $B_2(x,u)$
(equivalently, nearest to $y$ in $B_2(y,v)$).

$x_1$ denotes the common vertex of $B_1(x,u)$ and $B_2(y,v)$ nearest to $x$ in $B_1(x,u)$.

$y_1$ denotes the common vertex of $B_1(y,v)$ and $B_2(x,u)$ nearest to $y$ in $B_1(y,v)$.

$w_1$ denotes the common vertex of $B_2(x,u)$ and $B_2(y,v)$ nearest to $u$ in $B_2(x,u)$
(equivalently, nearest to $v$ in $B_2(y,v)$).

$u_1$ denotes the common vertex of $B_1(x,u)$ and $B_2(y,v)$ nearest to $u$ in $B_1(x,u)$.

$v_1$ denotes the common vertex of $B_1(y,v)$ and $B_2(x,u)$ nearest to $v$ in $B_1(y,v)$.

\noindent
In $G'$ we define $z'_1,x'_1,y'_1$ in the same way. However, $w'_1,u'_1,v'_1$ are defined
differently, by swapping the indices 1 and 2.

$w'_1$ denotes the common vertex of $B_1(x',u')$ and $B_1(y',v')$ nearest to $u'$ in $B_1(x',u')$
(equivalently, nearest to $v'$ in $B_1(y',v')$).

$u'_1$ denotes the common vertex of $B_2(x',u')$ and $B_1(y',v')$ nearest to $u'$ in $B_2(x',u')$.

$v'_1$ denotes the common vertex of $B_2(y',v')$ and $B_1(x',u')$ nearest to $v'$ in $B_2(y',v')$.
\end{definition}

\noindent
Note that some (and even all) of these six vertices can coincide.

\begin{lemma}{\assfont[Assumption \ref{ass:intersect}]}
\begin{bfenumerate}
\item
$B_1(x,u)[x_1,u_1]=B_2(y,v)[x_1,u_1]$ and the paths $B_1(x,u)$ and $B_2(y,v)$
have no other common point.
\item
$B_1(y,v)[y_1,v_1]=B_2(x,u)[y_1,v_1]$ and the paths $B_1(y,v)$ and $B_2(x,u)$
have no other common point.\noproof
\end{bfenumerate}
\end{lemma}

\begin{lemma}
\mbox{}
\begin{bfenumerate}
\item
$B_1(x',u')[x'_1,v'_1]=B_2(y',v')[x'_1,v'_1]$ and the paths $B_1(x',u')$ and $B_2(y',v')$
have no other common point.
\item
$B_1(y',v')[y'_1,u'_1]=B_2(x',u')[y'_1,u'_1]$ and the paths $B_1(y',v')$ and $B_2(x',u')$
have no other common point.\noproof
\end{bfenumerate}
\end{lemma}

We have proved the existence of two boudaries of $\calS_0(x,u)$
and $\calS_0(y,v)$ and fixed notation for them. We now extend it to other vertex pairs.

\begin{definition}\label{def:abbound}\rm
Let $C$ be an $X$- or $H$-configuration in a triconnected planar graph $G$.
Let $s$ and $t$ lie on a path $Q\in\calS_0(x,u)$, with $s$ being nearer to $x$.
Given $P\in\calS_0(s,t)$, let $\bar P$ denote the cycle $Q[x,s]PQ[t,u]wz$.
We call $P$ a \emph{boundary} of $\calS_0(s,t)$ if all $S_0(s,t)$ lie within the same region
bounded by~$\bar P$.
\end{definition}

\noindent
This definition does not depend on the choice of $Q$ nor on an embedding of $G$.
Similarly to Lemma \ref{lem:twobounds}, $\calS_0(s,t)$ has two boundaries, which
coincide if $\calS_0(s,t)$ is a singleton.

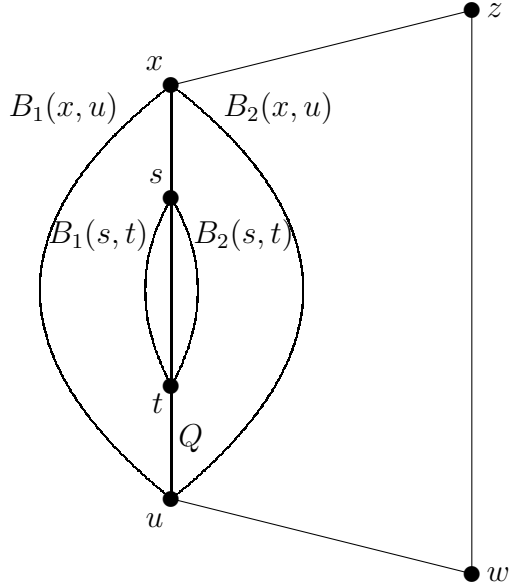
\begin{figure}[htbp]
\centerline{
\unitlength=1.00mm
\special{em:linewidth 0.4pt}
\linethickness{0.4pt}
\begin{picture}(92.00,83.00)
\put(50.00,72.00){\circle*{2.00}}
\put(50.00,57.00){\circle*{2.00}}
\put(50.00,32.00){\circle*{2.00}}
\put(50.00,17.00){\circle*{2.00}}
\bezier{220}(50.00,17.00)(50.00,44.00)(50.00,72.00)
\bezier{116}(50.00,57.00)(43.00,45.00)(50.00,32.00)
\bezier{116}(50.00,57.00)(57.00,45.00)(50.00,32.00)
\bezier{356}(50.00,72.00)(15.00,45.00)(50.00,17.00)
\bezier{356}(50.00,72.00)(85.00,45.00)(50.00,17.00)
\emline{50.00}{72.00}{1}{90.00}{82.00}{2}
\emline{90.00}{82.00}{3}{90.00}{7.00}{4}
\emline{90.00}{7.00}{5}{50.00}{17.00}{6}
\put(90.00,7.00){\circle*{2.00}}
\put(90.00,82.00){\circle*{2.00}}
\put(43.00,69.00){\makebox(0,0)[rc]{$B_1(x,u)$}}
\put(57.00,69.00){\makebox(0,0)[lc]{$B_2(x,u)$}}
\put(47.00,52.00){\makebox(0,0)[rc]{$B_1(s,t)$}}
\put(53.00,52.00){\makebox(0,0)[lc]{$B_2(s,t)$}}
\put(51.00,25.00){\makebox(0,0)[lc]{$Q$}}
\put(49.00,59.00){\makebox(0,0)[rb]{$s$}}
\put(49.00,31.00){\makebox(0,0)[rt]{$t$}}
\put(49.00,74.00){\makebox(0,0)[rb]{$x$}}
\put(49.00,15.00){\makebox(0,0)[rt]{$u$}}
\put(92.00,82.00){\makebox(0,0)[lc]{$z$}}
\put(92.00,7.00){\makebox(0,0)[lc]{$w$}}
\end{picture}
}
\caption{Two boundaries of $\calS_0(s,t)$.}%
\label{fig:5}
\end{figure}

\begin{convention}\rm
The two boundaries of $\calS_0(s,t)$ will be denoted by
$B_1(s,t)$ and $B_2(s,t)$ so that $B_1(s,t)$ is in the same region of $\bar B_2(s,t)$
together with $B_1(x,u)$, see Figure \ref{fig:5}.
As was said, similar definitions and notation will be used for the pair $yv$
in place of $xy$ and, furthermore, for the graph~$G'$.
\end{convention}

\begin{lemma}{\assfont[Assumption \ref{ass:intersect}]}\label{lem:Szw1}
$B_1(z_1,w_1)=B_2(y,v)[z_1,w_1]$ and hence contains $x_1$ and $u_1$.
$B_2(z_1,w_1)=B_2(x,u)[z_1,w_1]$ and hence contains $y_1$ and $v_1$.\noproof
\end{lemma}

\subsection{Forcing a shape of $\calS_0(z'_1,w'_1)$}

Despite some asymmetry between $G$ and $G'$ in the geometric Definition \ref{def:C1},
tuples of vertices $z_1,x_1,y_1,w_1,u_1,v_1$ and $z'_1,x'_1,y'_1,w'_1,u'_1,v'_1$
actually admit the same logical definition. 
In fact, the logical identity of these geometrically different things
is a key to Spoiler's strategy.
We state this logical identity in terms of the \EF\/ game.

\begin{lemma}{\assfont[Assumptions \ref{ass:eqdist} and \ref{ass:intersect}]}\label{lem:zetal1}
If Spoiler pebbles $z_1$, then Duplicator is forced to respond with $z'_1$
because otherwise Spoiler wins with 9 pebbles in $2\log n+3$ extra moves.
The same holds true for $x_1,w_1,u_1,y_1,v_1$.
\end{lemma}

\begin{proof}
Call a vertex $e$ an \emph{$x$-entrance} if it is the intersection point of some
$P\in\calS_0(x,u)$ with $B_2(y,v)$ nearest to $x$ in $P$. Note that $e$ is an $x$-entrance
iff
\begin{enumerate}
\item
$e\in S_0(x,u)\cap S_0(y,v)$ and
\item
there is an $x$-$e$-path of length $d_0(x,e)$ avoiding $S_0(y,v)$.
\end{enumerate}
It easily follows from Lemmas \ref{lem:d0respect} and \ref{lem:srespect} that
Duplicator is forced to respect the property of being an $x$- or $x'$-entrance:
Once $e$ and $e'$ are pebbled so that $e$ is an $x$-entrance but $e'$ is not an
$x'$-entrance or vice versa, Spoiler wins with 9 pebbles within the next $2\log n+2$
moves. For example, if Condition 2 is violated in $G'$, then we have
$d(x',e')\ne d(x,e)$ either in $G\setminus(C\setminus\{x\})$ and
$G'\setminus(C'\setminus\{x'\})$ or in $G\setminus(C\cup S_0(y,v)\setminus\{x,e\})$
and $G'\setminus(C'\cup S_0(y',v')\setminus\{x',e'\})$.

Geometrically, all $x$-entrances lie on $B_2(y,v)$. From this it
is easy to see that $z_1$ is the $x$-entrance $e$ minimizing $d_0(e,y)$
and $x_1$ is the $x$-entrance $e$ maximizing $d_0(e,y)$.
Thus, if Duplicator responds to $z_1$ (resp.\ $x_1$) not with $z'_1$ (resp.\ $x'_1$),
then in the next round Spoiler does pebble $z'_1$ (resp.\ $x'_1$), which
forces Duplicator to violate either the equality relation or the condition of being an entrance.

For $w_1,u_1,y_1,v_1$ the lemma is proved by a symmetric argument.
\end{proof}

For our further analysis we make the following assumption. If it is not true,
Spoiler wins with 9 pebbles in $2\log n+5$ moves by Lemmas \ref{lem:zetal1} and~\ref{lem:d0respect}.

\begin{assumption}\label{ass:d0equal}\rm
The $d_0$-distances between the vertices $x'$, $y'$, $u'$, $v'$, $z'_1$, $w'_1$, $x_1'$, $y'_1$,
$u'_1$, and $v'_1$ in $G'$ are equal to the corresponding $d_0$-distances in~$G$.
\end{assumption}

This assumption enables us to determine the boundaries of $\calS_0(z'_1,w'_1)$.
Note that no analog of Lemma \ref{lem:Szw1} for $G'$ (i.e., for the twisted case) follows just
from the definitions.

\begin{lemma}{\assfont[Assumptions \ref{ass:intersect} and \ref{ass:d0equal}]}\label{lem:Szw11}
\begin{bfenumerate}
\item
$z'_1$ and $w'_1$ belong to a path in $\calS_0(x',u')$.
\item
Define two $z'_1$-$w'_1$-paths $B_1$ and $B_2$ by $B_1=B_2(y',v')[z'_1,x'_1]B_1(x',u')[x'_1,w'_1]$
and $B_2=B_2(x',u')[z'_1,u'_1]B_1(y',v')[u'_1,w'_1]$. Then
$B_1(z'_1,w'_1)=B_1$ and hence contains $x'_1$ and $v'_1$. Furthermore,
$B_2(z'_1,w'_1)=B_2(x,u)$ and hence contains $y'_1$ and~$u'_1$.
\end{bfenumerate}
\end{lemma}

\begin{proof}
We have $d_0(x',u')=d_0(x',z'_1)+d_0(z'_1,w'_1)+d_0(w'_1,u')$
because this is so in $G$. This equality means that $z'_1$ and $w'_1$
lie on a $d_0$-geodesic between $x'$ and $u'$.
Similarly, $z'_1$ and $w'_1$ lie on a $d_0$-geodesics between $y'$ and $v'$.
It follows that $S_0(z'_1,w'_1)\subset S_0(x',u')\cap S_0(y',v')$.
This inclusion implies that $S_0(z'_1,w'_1)$ is inside of regions bounded by
$\bar B_1$ and $\bar B_2$, the cycles defined as in Definition \ref{def:abbound}.
Therefore, $B_1$ and $B_2$ are boundaries of $\calS_0(z'_1,w'_1)$ provided
they are in $\calS_0(z'_1,w'_1)$. The latter condition is true indeed.
Let us show this, say, for $B_1$. We have 
$d_0(z'_1,u'_1)=d_0(z'_1,x'_1)+d_0(x'_1,u'_1)$ because this is so in $G$.
Hence $x'_1$ lies on a $d_0$-geodesic from $z'_1$ to $w'_1$, which gives us what
we need because $B_1$ consists of two $d_0$-geodesics, from $z'_1$ to $x'_1$ and
from $x'_1$ to $w'_1$.
\end{proof}

\begin{lemma}\label{lem:blockpath}
Let $s$ and $t$ belong to a path in $\calS_0(x,u)$.
Then $G[S_0(s,t)]$ has simple cut-block relation 
(as defined in Section \ref{ss:graphs})
and its block-tree is a path.
\end{lemma}

\begin{proof}
Assume, to the contrary, that three blocks of $G[S_0(s,t)]$ share the same
cutpoint $c$. Let $B$ be one of these blocks containing neither $s$ nor $t$.
Let $a$ be a vertex in $B$ different from $c$. As $a\in S_0(s,t)$, there is
a path $P\in\calS_0(s,t)$ going through $a$, which should cross $c$ twice,
once on the segment $P[s,a]$ and once again on the segment $P[a,t]$.
This gives us a contradiction.

The assumption that a block of $G[S_0(s,t)]$ has three neighbors in the block-tree
leads to a contradiction in a similar way.
\end{proof}

Note that the cutpoints of $G[S_0(s,t)]$ are exactly the
common points of $B_1(a,b)$ and $B_2(a,b)$.

Lemma \ref{lem:blockpath} applies to $G[S_0(z_1,w_1)]$ and,
under Assumptions \ref{ass:intersect} and \ref{ass:d0equal}, also to $G'[S_0(z'_1,w'_1)]$.

\begin{lemma}\label{lem:cutrespect}
Let vertices $s,t\in V(G)$ and $s',t'\in V(G')$ be pebbled so that $d_0(s,t)=d_0(s',t')$.
Then Duplicator is forced to respect cutpoints of $G[S_0(s,t)]$ and $G'[S_0(s',t')]$:
Whenever Spoiler pebbles a cutpoint of $G[S_0(s,t)]$ or $G'[S_0(s',t')]$,
Duplicator must respond with a cutpoint in the other graph because otherwise Spoiler
wins with 12 pebbles in less than $2\log n+2$ extra moves.
\end{lemma}

\begin{proof}
Spoiler will play in $G[S_0(s,t)]$ and $G'[S_0(s',t')]$ and, by Lemma \ref{lem:srespect},
we will assume that Duplicator respects this restriction.
Let Spoiler pebble, say, a cutpoint $c$ of $G[S_0(s,t)]$.
Let $c'$ be Duplicator's response in $G'[S_0(s',t')]$. If $c'$ is not a cutpoint,
Spoiler restricts further play to
graphs $G[S_0(s,t)]-c$ and $G'[S_0(s',t')]-c'$ 
and wins fast because the former graph is connected while the latter is not.
\end{proof}

By Assumption \ref{ass:d0equal}, by Lemma \ref{lem:cutrespect} applied to $s=z_1,\,t=w_1$, 
and by Lemma \ref{lem:d0respect}, 
we see that, unless Spoiler wins with 12 pebbles in $2\log n+3$ moves,
the following condition is true.

\begin{assumption}\label{ass:cuts}\rm
There is a one-to-one correpondence between the cutpoints of $G[S_0(z_1,w_1)]$ and 
$G'[S_0(z'_1,w'_1)]$ such that, if $c$ is a cutpoint of $G[S_0(z_1,w_1)]$,
then $G'[S_0(z'_1,w'_1)]$ has a unique vertex $c'$ with $d_0(z'_1,c')=d_0(z_1,c)$,
which is a cutpoint in this graph, and vice versa.
\end{assumption}

Note that the correspondence between the cutpoints determines a one-to-one correspondence
between the blocks of $G[S_0(z_1,w_1)]$ and $G'[S_0(z'_1,w'_1)]$.

\subsection{A basic stratagem of Spoiler}

\begin{figure}[htbp]
\centerline{
\unitlength=1.00mm
\special{em:linewidth 0.4pt}
\linethickness{0.4pt}
\begin{picture}(148.00,64.00)
\put(43.00,62.00){\circle*{2.00}}
\put(43.00,7.00){\circle*{2.00}}
\bezier{356}(43.00,62.00)(8.00,35.00)(43.00,7.00)
\bezier{356}(43.00,62.00)(78.00,35.00)(43.00,7.00)
\put(36.00,59.00){\makebox(0,0)[rc]{$B_1(s,t)$}}
\put(50.00,59.00){\makebox(0,0)[lc]{$B_2(s,t)$}}
\put(43.00,64.00){\makebox(0,0)[cb]{$s$}}
\put(43.00,5.00){\makebox(0,0)[ct]{$t$}}
\put(33.00,53.00){\circle*{2.00}}
\put(113.00,62.00){\circle*{2.00}}
\put(113.00,7.00){\circle*{2.00}}
\bezier{356}(113.00,62.00)(78.00,35.00)(113.00,7.00)
\bezier{356}(113.00,62.00)(148.00,35.00)(113.00,7.00)
\put(106.00,59.00){\makebox(0,0)[rc]{$B_1(s',t')$}}
\put(120.00,59.00){\makebox(0,0)[lc]{$B_2(s',t')$}}
\put(113.00,64.00){\makebox(0,0)[cb]{$s'$}}
\put(113.00,5.00){\makebox(0,0)[ct]{$t'$}}
\put(103.00,53.00){\circle*{2.00}}
\put(31.00,53.00){\makebox(0,0)[rb]{$a$}}
\put(101.00,53.00){\makebox(0,0)[rb]{$a'$}}
\put(34.00,15.00){\circle*{2.00}}
\put(121.00,14.00){\circle*{2.00}}
\put(123.00,13.00){\makebox(0,0)[lt]{$b'$}}
\put(32.00,13.00){\makebox(0,0)[rt]{$b$}}
\end{picture}
}
\caption{An initial position in Lemma \protect\ref{lem:stratagem}.}%
\label{fig:6}
\end{figure}
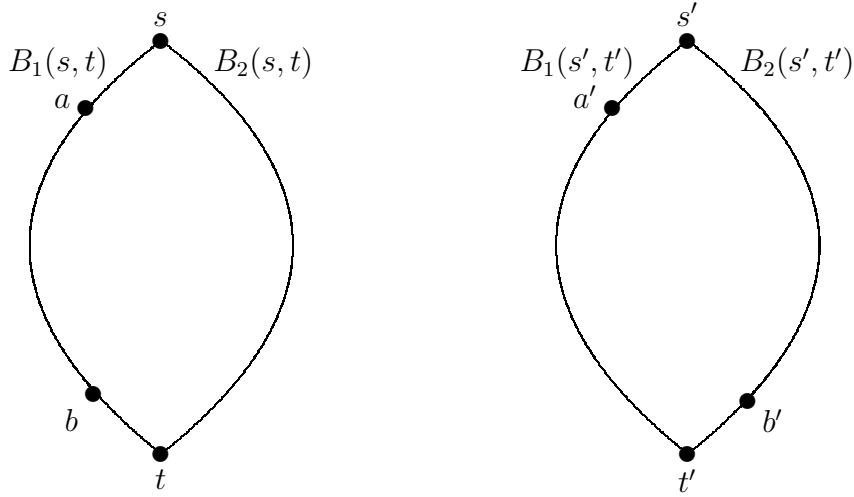

\begin{lemma}\label{lem:stratagem}{\assfont[Assumption \ref{ass:eqdist}]}
Let pairwise distinct vertices $s,t,a,b\in V(G)$ and $s',t',a',b'\in V(G')$ be pebbled so that
all corresponding $d_0$-distances between all pebbled vertices in $G$ and $G'$ are equal.
Let $s$ and $t$ lie on a $d_0$-geodesic from $x$ to $u$ (hence $s'$ and $t'$ lie on a 
$d_0$-geodesic from $x'$ to $u'$). 
Suppose that $G[S_0(s,t)]$ is biconnected (i.e., $B_1(s,t)$ and $B_2(s,t)$ have no common
inner point). Furthermore, assume that $a$ and $b$ lie on the same
boundary of $\calS_0(s,t)$ while $a'$ and $b'$ lie on different
boundaries of $\calS_0(s',t')$, see Figure \ref{fig:6}. 
Then Spoiler wins with 13 pebbles in $3\log n+5$ moves.
\end{lemma}

\begin{proof}
We will suppose that $G'[S_0(s',t')]$ is biconnected too
because otherwise Spoiler has a fast win by Lemma \ref{lem:cutrespect}.
Without loss of generality, suppose that both $a$ and $b$ 
lie on $B_1(s,t)$, $a$ is nearer to $s$, $a'$ lies on $B_1(s',t')$, 
and $b'$ on $B_2(s',t')$.
Set $a_1=a$ and $a'_1=a'$ and define a sequence of $a_i$'s and a sequence of $a'_i$'s
by an inductive rule. Suppose that vertices $a_i$ and $a'_i$ are already defined so that, 
if $i$ is odd, $a_i$ and $a'_i$ are inner vertices of, respectively, $B_1(s,t)$ and
$B_1(s',t')$ and, if $i$ is even, $a_i$ and $a'_i$ are inner vertices of, respectively, $B_2(s,t)$ and
$B_2(s',t')$. 

Given $p$ and $q$ being inner points of a path in $\calS_0(s,t)$, we say that $S_0(p,q)$ \emph{blocks}
$\calS_0(s,t)$ if $G[S_0(s,t)]\setminus S_0(p,q)$ is disconnected.
Call $a_i$ \emph{blocking} if $S_0(a_i,b)$ blocks $\calS_0(s,t)$. 
The similar definition will be used for $a'_i$. Note that even $a_i$'s and
odd $a'_i$'s are always blocking.
Call the pair $a_i,a'_i$ \emph{distinguishing} if  only one of the two vertices is blocking
or if $d(a_i,b)\ne d(a'_i,b')$.
Note that $a_i,a'_i$ is necessary distinguishing if $a_i$ and $b$ are adjacent.

In the case that the pair $a_i,a'_i$ is non-distinguishing, we define $a_{i+1}$ and
$a'_{i+1}$ as follows. Let $i$ be odd. The condition that $a_i$ is blocking means that
$B_2(a_i,b)$ touches $B_2(s,t)$. Geometrically, we define $a_{i+1}$ to be the common point
of $B_2(a_i,b)$ and $B_2(s,t)$ nearest to $a_i$ and $s$ in these paths, see Figure~\ref{fig:7}.

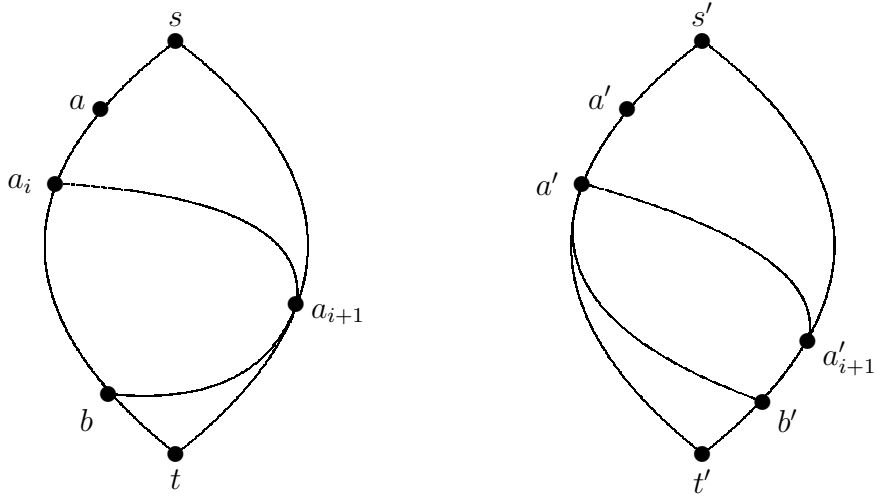
\begin{figure}[htbp]
\centerline{
\unitlength=1.00mm
\special{em:linewidth 0.4pt}
\linethickness{0.4pt}
\begin{picture}(148.00,65.00)
\put(43.00,63.00){\circle*{2.00}}
\put(43.00,8.00){\circle*{2.00}}
\bezier{356}(43.00,63.00)(8.00,36.00)(43.00,8.00)
\bezier{356}(43.00,63.00)(78.00,36.00)(43.00,8.00)
\put(43.00,65.00){\makebox(0,0)[cb]{$s$}}
\put(43.00,6.00){\makebox(0,0)[ct]{$t$}}
\put(33.00,54.00){\circle*{2.00}}
\put(113.00,63.00){\circle*{2.00}}
\put(113.00,8.00){\circle*{2.00}}
\bezier{356}(113.00,63.00)(78.00,36.00)(113.00,8.00)
\bezier{356}(113.00,63.00)(148.00,36.00)(113.00,8.00)
\put(113.00,65.00){\makebox(0,0)[cb]{$s'$}}
\put(113.00,6.00){\makebox(0,0)[ct]{$t'$}}
\put(103.00,54.00){\circle*{2.00}}
\put(31.00,54.00){\makebox(0,0)[rb]{$a$}}
\put(101.00,54.00){\makebox(0,0)[rb]{$a'$}}
\put(34.00,16.00){\circle*{2.00}}
\put(121.00,15.00){\circle*{2.00}}
\put(123.00,14.00){\makebox(0,0)[lt]{$b'$}}
\put(32.00,14.00){\makebox(0,0)[rt]{$b$}}
\put(127.00,23.00){\circle*{2.00}}
\put(97.00,44.00){\circle*{2.00}}
\put(27.00,44.00){\circle*{2.00}}
\put(59.00,28.00){\circle*{2.00}}
\bezier{140}(59.00,28.00)(54.00,14.00)(34.00,16.00)
\bezier{192}(59.00,28.00)(61.00,42.00)(27.00,44.00)
\bezier{188}(97.00,44.00)(130.00,35.00)(127.00,23.00)
\bezier{208}(121.00,15.00)(90.00,26.00)(97.00,44.00)
\put(94.00,44.00){\makebox(0,0)[rc]{$a'$}}
\put(24.00,44.00){\makebox(0,0)[rc]{$a_i$}}
\put(61.00,28.00){\makebox(0,0)[lt]{$a_{i+1}$}}
\put(129.00,23.00){\makebox(0,0)[lt]{$a'_{i+1}$}}
\end{picture}
}
\caption{Proof of Lemma \protect\ref{lem:stratagem}.}%
\label{fig:7}
\end{figure}

The $a_{i+1}$
admits also a logical definition: it is a (unique) vertex $r$ such that
\begin{itemize}
\item
$r\in S_0(a_i,b)$,
\item
$S_0(a_i,r)$ blocks $\calS_0(s,t)$,
\item
$d_0(a_i,r)$ (equivalently, $d_0(s,r)$) is minimum possible.
\end{itemize}
$a'_{i+1}$ is defined, both geometrically and logically, in exactly the same way.
If $i$ is even, $a_{i+1}$ and $a'_{i+1}$ are defined 
geometrically in a similar way and logically in literally the same way.

If $i$ is odd, then $a_{i+1}$ lies on a $d_0$-geodesic strictly between $a_i$ and $b$.
If $i$ is even, this is also true except the case that perhaps $a_{i+1}=b$.
Note that in the latter case $a'_{i+1}$ still lies strictly between $a'_i$ and $b'$ on a $d_0$-geodesic
and hence $a'_{i+1}\ne b'$. 

Thus, $d_0(a_{i},b)$ each time decreases and can eventually become 1 or 0.
In both these cases the pair $a_i,a'_i$ becomes distinguishing.

The logical definition of $a_{i+1}$ and $a'_{i+1}$ shows that, if the vertices $s,t,a_i,b$
and $s',t',a'_i,b'$ are pebbled in $G$ and $G'$ respectively and Spoiler pebbles $a_{i+1}$
(resp.\ $a'_{i+1}$), then Duplicator is forced to respond with $a'_{i+1}$ (resp.\ $a_{i+1}$) 
because otherwise Spoiler wins
with 12 pebbles in $\log d_0(s,t)+\log d_0(a,b)+3\le2\log n+3$ extra moves similarly to the proof of 
Lemma~\ref{lem:zetal1}.

We are now prepared to describe Spoiler's strategy.
Let $m$ denote the largest $i$ for which $a_i$ and $a'_i$ are defined.
Spoiler starts with pebbling $a_m$. If Duplicator responds not with $a'_m$,
Spoiler applies the Generalized Halving Strategy and, by Lemma \ref{lem:exthalving}, 
wins with 12 pebbles in $\log m+2\log n+4\le3\log n+4$ moves.

We hence assume that Duplicator pebbles $a'_m$.
Recall that the pair $a_m,a'_m$ is distinguishing. 
Let the $d_0$-distance be preserved; otherwise we are done by Lemma \ref{lem:d0respect}.
Thus, $G$ and $G'$ disagree with respect to the blocking property.
Suppose, for instance, that
$a'_m$ is blocking but $a_m$ is not. Based on Lemma \ref{lem:srespect}, Spoiler
restricts further play to $G[S_0(s,t)]\setminus S_0(a_m,b)$ and $G'[S_0(s',t')]\setminus S_0(a'_m,b')$
and wins with 13 pebbles in $\log d_0(s,t)+1$ moves (however, yet other $\log d_0(s,t)$ 
moves are needed if Duplicator decides to move outside these subgraphs).
\end{proof}

\subsection{The case of $\calS_0(x,u)$ and $\calS_0(y,v)$ not having the strong
intersection property}\label{ss:notstrong}

In Subsection \ref{ss:notint} we analized the case that $\calS_0(x,u)$ and $\calS_0(y,v)$
do not have the intersection property and assumed for the further analysis that this property
is true. The next case, that we consider now, is that
$\calS_0(x,u)$ and $\calS_0(y,v)$ do not have 
the strong intersection property or, equivalently, that $B_1(x,u)$ and $B_1(y,v)$
have no common point. 

In this case $x_1,y_1,u_1,v_1$ all belong to the same block of $G[S_0(z_1,w_1)]$.
Denote the cutpoints belonging to this block by $s$ and $t$ (or, if this block
is an endpoint of the block-tree of $G[S_0(z_1,w_1)]$,
then $s=z_1$ or $t=w_1$). By Assumptions \ref{ass:d0equal} and \ref{ass:cuts},
$x'_1,y'_1,u'_1,v'_1$ all belong to the block of $G'[S_0(z'_1,w'_1)]$
which is cut by the cutpoints $s'$ and $t'$ corresponding to $s$ and $t$.
In the first six rounds Spoiler pebbles $x_1,y_1,u_1,v_1$ and $s,t$.
Suppose that Duplicator responds with $x'_1,y'_1,u'_1,v'_1$ and $s',t'$
respectively. If not, then Spoiler wins with 12 pebbles in $2\log n+3$ extra moves 
by Lemmas \ref{lem:zetal1} and~\ref{lem:d0respect}.

Assume that Duplicator has not lost so far, in particular, any two vertices of $x_1,y_1,u_1,v_1$ 
are equal iff such are the corresponding two of $x'_1,y'_1,u'_1,v'_1$.
In this case all $x_1,y_1,u_1,v_1$ (and $x'_1,y'_1,u'_1,v'_1$) are pairwise disjoint.
Indeed, $\{x_1,u_1\}$ and $\{y_1,v_1\}$ are disjoint because they lie on $B_1(x,u)$ and $B_1(y,v)$,
respectively. If $x_1=u_1$, then it is easy to see from Lemma \ref{lem:Szw11} 
that $x'_1=u'_1$ is a cutpoint of $G[S_0(z_1,w_1)]$
and hence $x_1=u_1$ should be a cutpoint of $G[S_0(z_1,w_1)]$, which is impossible.
Thus, $x_1\ne u_1$ and, similarly, $y_1\ne v_1$.

Suppose that $\{x_1,u_1\}$ and $\{s,t\}$ are disjoint and, hence, so are
$\{x'_1,u'_1\}$ and $\{s',t'\}$. In this case we are in the conditions
of Lemma \ref{lem:stratagem} with $a=x_1,b=u_1$ and hence Spoiler wins with 13 pebbles in $3\log n+5$ 
extra moves (the pebbles on $y_1$ and $v_1$ can be reused). 
The case of disjoint $\{y_1,v_1\}$ and $\{s,t\}$ is symmetric.

It remains to consider the case that both $\{x_1,u_1\}$ and $\{y_1,v_1\}$ 
intersect $\{s,t\}$. Let, say, $x_1=s$ and $v_1=t$. Set $a=y_1$ and $b=u_1$.
We are in the conditions of Lemma \ref{lem:stratagem} with $G$ and $G'$ interchanged
and again Spoiler has an efficient win.

Thus, in any case Spoiler wins with 13 pebbles in at most $3\log n+11$ moves.

\subsection{The case of $\calS_0(x,u)$ and $\calS_0(y,v)$ with the strong
intersection property}

Another formulation of the case treated here is that $B_1(x,u)$ and $B_1(y,v)$ touch.
Notice the principal distinction between the cases considered here and in Section 
\ref{ss:notstrong}. In Section \ref{ss:notstrong}, Spoiler was able, staying all
the time in $S_0(x,u)\cup S_0(y,v)$, to exhibit
the difference between the case that $a,b\in B_1(x,u)$ and the case that
$a\in B_1(x,u),\,b\in B_1(y,v)$. Now this is impossible in principle; say, if
$a$ and $b$ are separated by a vertex $c$ at which $B_1(x,u)$ and $B_1(y,v)$
touch, then Spoiler has to go outside $S_0(x,u)\cup S_0(y,v)$.

Denote
$$
H=G[S_0(x,u)\cup S_0(y,v)]\mbox{\ \ and\ \ }
H'=G[S_0(x',u')\cup S_0(y',v')].
$$
The play will much depend on the structure of graph~$H$.

\begin{figure}[htbp]
\centerline{
\unitlength=1.00mm
\special{em:linewidth 0.4pt}
\linethickness{0.4pt}
\begin{picture}(90.00,97.00)
\put(30.00,96.00){\circle*{2.00}}
\put(80.00,96.00){\circle*{2.00}}
\put(55.00,81.00){\circle*{2.00}}
\bezier{84}(55.00,81.00)(48.00,73.00)(55.00,66.00)
\bezier{84}(55.00,66.00)(62.00,73.00)(55.00,81.00)
\bezier{84}(55.00,66.00)(48.00,58.00)(55.00,51.00)
\bezier{84}(55.00,51.00)(62.00,58.00)(55.00,66.00)
\bezier{84}(55.00,51.00)(48.00,43.00)(55.00,36.00)
\bezier{84}(55.00,36.00)(62.00,43.00)(55.00,51.00)
\bezier{84}(55.00,36.00)(48.00,28.00)(55.00,21.00)
\bezier{84}(55.00,21.00)(62.00,28.00)(55.00,36.00)
\bezier{128}(55.00,81.00)(43.00,96.00)(30.00,96.00)
\bezier{128}(55.00,81.00)(67.00,96.00)(80.00,96.00)
\bezier{184}(80.00,96.00)(61.00,71.00)(58.00,57.00)
\put(58.00,59.00){\circle*{2.00}}
\bezier{132}(30.00,96.00)(46.00,85.00)(52.00,73.00)
\put(52.00,73.00){\circle*{2.00}}
\put(55.00,51.00){\circle*{2.00}}
\put(55.00,36.00){\circle*{2.00}}
\put(55.00,21.00){\circle*{2.00}}
\put(30.00,6.00){\circle*{2.00}}
\put(80.00,6.00){\circle*{2.00}}
\bezier{140}(80.00,6.00)(62.00,6.00)(55.00,21.00)
\bezier{140}(30.00,6.00)(48.00,6.00)(55.00,21.00)
\bezier{144}(30.00,6.00)(49.00,16.00)(52.00,30.00)
\put(52.00,30.00){\circle*{2.00}}
\bezier{136}(80.00,6.00)(59.00,20.00)(58.00,29.00)
\put(58.00,29.00){\circle*{2.00}}
\put(10.00,51.00){\dashbox{2.00}(80.00,0.00)[cc]{}}
\put(10.00,36.00){\dashbox{2.00}(80.00,0.00)[cc]{}}
\put(15.00,53.00){\makebox(0,0)[cb]{$H_0$}}
\put(15.00,43.00){\makebox(0,0)[cc]{$H_1$}}
\put(15.00,34.00){\makebox(0,0)[ct]{$H_2$}}
\put(27.00,96.00){\makebox(0,0)[rc]{$x$}}
\put(83.00,96.00){\makebox(0,0)[lc]{$y$}}
\put(83.00,6.00){\makebox(0,0)[lc]{$v$}}
\put(27.00,6.00){\makebox(0,0)[rc]{$u$}}
\put(55.00,17.00){\makebox(0,0)[ct]{$w_1$}}
\put(55.00,84.00){\makebox(0,0)[cb]{$z_1$}}
\put(49.00,73.00){\makebox(0,0)[rt]{$x_1$}}
\put(60.00,59.00){\makebox(0,0)[lt]{$y_1$}}
\put(52.00,52.00){\makebox(0,0)[rb]{$e_1$}}
\put(52.00,37.00){\makebox(0,0)[rb]{$e_2$}}
\put(49.00,29.00){\makebox(0,0)[rc]{$u_1$}}
\put(61.00,28.00){\makebox(0,0)[lc]{$v_1$}}
\end{picture}
}
\caption{Splitting $H$ by essential cutpoints.}%
\label{fig:8}
\end{figure}
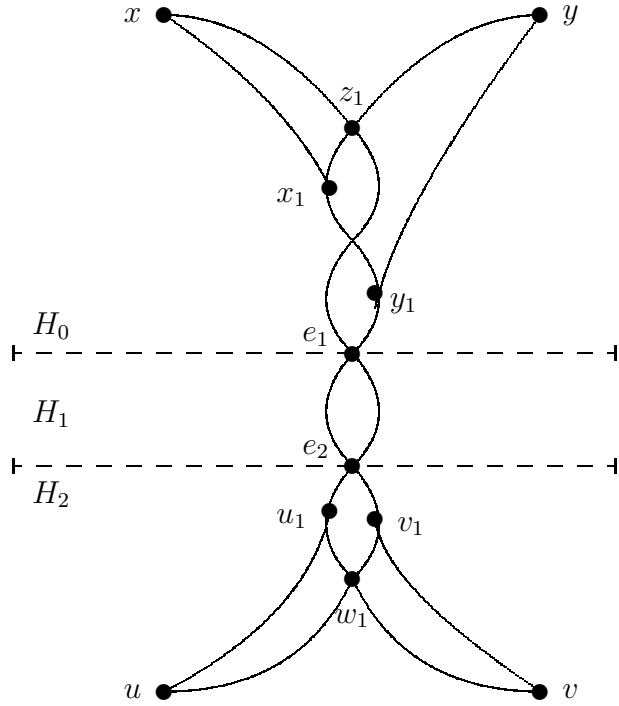

We introduce some terminology for $H$ which will be used as well for $H'$.
We call a cutpoint $c$ of $H$ \emph{essential} if $c$ is also a cutpoint of $G[S_0(z_1,w_1)]$,
or $c=x_1=y_1=z_1$, or $c=u_1=v_1=w_1$.
In this subsection we suppose that $\calS_0(x,u)$ and $\calS_0(y,v)$ have the strong
intersection property. This implies that $H$ has at least one essential cutpoint.
Let $e_1,\ldots,e_l$ be the essential cutpoints of $H$ listed so that $d_0(z_1,e_i)$ increases.
We split $K$ by $e_1,\ldots,e_l$ into subgraphs
$
H_0,\ldots,H_l
$
so that
\begin{eqnarray*}
&&H_0=G[S_0(x,e_1)\cup S_0(y,e_1)],\quad
H_l=G[S_0(e_l,u)\cup S_0(e_l,v)],\\
&&H_i=G[S_0(e_{i-1},e_i)]\mbox{\ if\ }0<i<l.
\end{eqnarray*}
Thus, any $H_i$ for $0<i<l$ is a block of $G[S_0(z_1,w_1)]$, see Figure~\ref{fig:8}.

Note that $e_1$ is definable as the vertex $c$ such that $c=z_1$ or $c$ is a cutpoint of 
$G[S_0(z_1,w_1)]$, $d_0(z_1,c)\ge\max\{d_0(z_1,x_1),d_0(z_1,y_1)\}$, and $d_0(z_1,c)$
is minimum possible. The $e_l$ admits a similar definition in terms of $u_1$, $v_1$,
and $w_1$. We hence state, as a consequence of Assumptions \ref{ass:d0equal} and \ref{ass:cuts},
a one-to-one corrrespondence between the $e_i$'s and the $e'_i$'s as well as between
the $H_i$'s and the $H'_i$'s.

\begin{assumption}\label{ass:essential}\rm
$H$ and $H'$ have the same number of essential cutpoints and
$d_0(z_1,e_i)=d_0(z'_1,e'_i)$ for all $i\le l$.
\end{assumption}

As was mentioned in the beginning of this section, playing inside $H$ and $H'$, Spoiler cannot
force Duplicator to respect the $x$-sides and $y$-sides of $H$ and $H'$.
However, for $H_0$ and $H'_0$ this is still possible.

\begin{figure}[htbp]
\centerline{
\unitlength=1.00mm
\special{em:linewidth 0.4pt}
\linethickness{0.4pt}
\begin{picture}(143.00,56.00)
\put(10.00,55.00){\circle*{2.00}}
\put(60.00,55.00){\circle*{2.00}}
\put(35.00,40.00){\circle*{2.00}}
\bezier{84}(35.00,40.00)(28.00,32.00)(35.00,25.00)
\bezier{84}(35.00,25.00)(42.00,32.00)(35.00,40.00)
\bezier{84}(35.00,25.00)(28.00,17.00)(35.00,10.00)
\bezier{84}(35.00,10.00)(42.00,17.00)(35.00,25.00)
\bezier{128}(35.00,40.00)(23.00,55.00)(10.00,55.00)
\bezier{128}(35.00,40.00)(47.00,55.00)(60.00,55.00)
\bezier{184}(60.00,55.00)(41.00,30.00)(38.00,16.00)
\put(38.00,18.00){\circle*{2.00}}
\bezier{132}(10.00,55.00)(26.00,44.00)(32.00,32.00)
\put(32.00,32.00){\circle*{2.00}}
\put(35.00,10.00){\circle*{2.00}}
\put(7.00,55.00){\makebox(0,0)[rc]{$x$}}
\put(63.00,55.00){\makebox(0,0)[lc]{$y$}}
\put(35.00,43.00){\makebox(0,0)[cb]{$z_1$}}
\put(29.00,31.00){\makebox(0,0)[ct]{$x_1$}}
\put(40.00,18.00){\makebox(0,0)[lt]{$y_1$}}
\put(12.00,32.00){\makebox(0,0)[cc]{$B_1(x,e_1)$}}
\put(90.00,55.00){\circle*{2.00}}
\put(140.00,55.00){\circle*{2.00}}
\put(115.00,40.00){\circle*{2.00}}
\bezier{84}(115.00,40.00)(108.00,32.00)(115.00,25.00)
\bezier{84}(115.00,25.00)(122.00,32.00)(115.00,40.00)
\bezier{84}(115.00,25.00)(108.00,17.00)(115.00,10.00)
\bezier{84}(115.00,10.00)(122.00,17.00)(115.00,25.00)
\bezier{128}(115.00,40.00)(103.00,55.00)(90.00,55.00)
\bezier{128}(115.00,40.00)(127.00,55.00)(140.00,55.00)
\put(115.00,10.00){\circle*{2.00}}
\put(87.00,55.00){\makebox(0,0)[rc]{$y'$}}
\put(143.00,55.00){\makebox(0,0)[lc]{$x'$}}
\put(115.00,43.00){\makebox(0,0)[cb]{$z'_1$}}
\put(92.00,32.00){\makebox(0,0)[cc]{$B_1(y',e'_1)$}}
\put(35.00,7.00){\makebox(0,0)[ct]{$e_1$}}
\put(115.00,7.00){\makebox(0,0)[ct]{$e'_1$}}
\bezier{180}(112.00,16.00)(109.00,28.00)(90.00,55.00)
\put(111.00,18.00){\circle*{2.00}}
\put(109.00,17.00){\makebox(0,0)[rb]{$y'_1$}}
\bezier{140}(118.00,29.00)(127.00,46.00)(140.00,55.00)
\put(118.00,30.00){\circle*{2.00}}
\put(120.00,30.00){\makebox(0,0)[lt]{$x'_1$}}
\put(99.00,42.00){\circle*{2.00}}
\put(96.00,43.00){\makebox(0,0)[rt]{$a'$}}
\put(24.00,43.00){\circle*{2.00}}
\put(21.00,43.00){\makebox(0,0)[rt]{$a$}}
\end{picture}
}
\caption{An initial position in Lemma \protect\ref{lem:siderespect}.}%
\label{fig:9}
\end{figure}
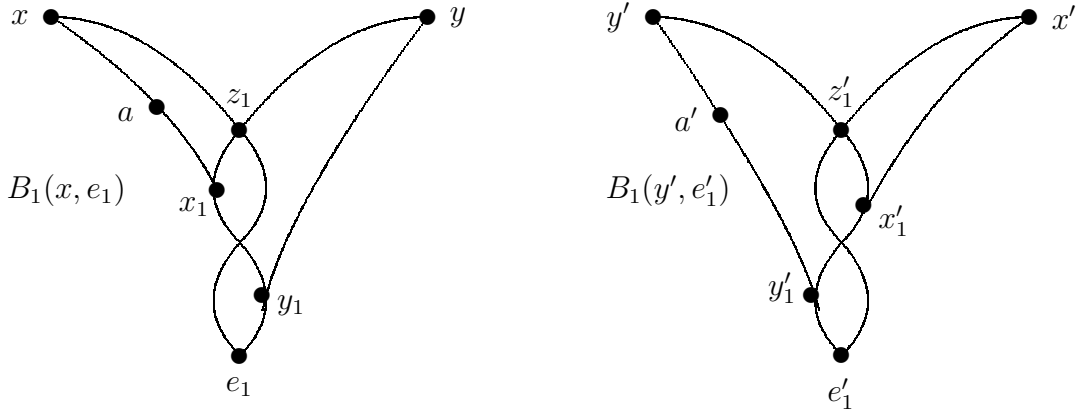

\begin{lemma}{\assfont[Assumptions \ref{ass:d0equal}, \ref{ass:cuts}, \ref{ass:essential}]}%
\label{lem:siderespect}
Let $a$ lie on $B_1(x,e_1)$ and $a'\ne e'_1$ lie on $B_1(y',e'_1)$, see Figure \ref{fig:9}.
Suppose that $a$ and $a'$ are under the same pebble.
Then Spoiler wins with 13 pebbles in $3\log n+8$ moves.
The same holds true with $x$ and $y$ interchanged.
Furthermore, the symmetric claim is true for $H_l$ and $H'_l$
(where the symmetry is with respect to the substitution $(xu)(yv)(e_1e_l)(B_1B_2)$).
\end{lemma}

\begin{proof}
Spoiler first pebbles $z_1$, $x_1$, and $e_1$. Suppose that Duplicator, to avoid
a fast loss, responds with $z'_1$, $x'_1$, and $e'_1$.
The case of $a\in\{x_1,e_1\}$ is trivial, we hence assume these vertices distinct.

\case 1{$a\in B_1(x,x_1)$.}
If $a'\in B_1(y',y'_1)$, we have $a\in S_0(x,e_1)$ while $a'\notin S_0(x',e'_1)$.
If $a'\in B_1(y'_1,e'_1)$, we have $a'\in S_0(z'_1,e'_1)$ while $a\notin S_0(z_1,e_1)$.
In both cases Spoiler wins by Lemma~\ref{lem:srespect}.

\case 2{$a\in B_1(x_1,e_1)$.}
It is possible that $a$ is a cutpoint of $G[S_0(z_1,e_1)]$.
It easily follows from Assumptions \ref{ass:d0equal} and \ref{ass:cuts}, that
$x_1$ and $x'_1$ as well as $y_1$ and $y'_1$ belong to corresponding blocks
of $G[S_0(z_1,w_1)]$ and $G'[S_0(z'_1,w'_1)]$. From this fact and from
the definition of $e_1$, we see that $a'$ cannot be a cutpoint of $G'[S_0(z'_1,e'_1)]$
and hence Spoiler wins by Lemma~\ref{lem:cutrespect}.

Assume now that $a$ is not a cutpoint. To make Spoiler's life harder, we also
assume that $d_0(z_1,a)=d_0(z'_1,a')$. This implies that $a$ and $a'$ belong to
corresponding blocks of $G[S_0(z_1,w_1)]$ and $G'[S_0(z'_1,w'_1)]$, say, $K$ and $K'$.
By the definition of $e_1$, $K$ should also contain $y_1$ or $x_1$
(the former is depicted in Figure \ref{fig:9})
and the same holds for $K'$. This allows Spoiler to apply the basic stratagem
given by Lemma~\ref{lem:stratagem}.
\end{proof}

For description of Spoiler's strategy, we need some further definitions.
Even if introduced for $G$, they will be used for both $G$ and $G'$.
We call an $a$-$b$-path $P$ \emph{external} if both $a$ and $b$ are in $H$ and
no inner vertex of $P$ is in $V(H)\cup\{z,w\}$. This definition is logical,
in the following sense.

\begin{lemma}\label{lem:extrespect}
Let $a,b\in V(G)$ and $a',b'\in V(G')$ be pebbled so that
$a$ and $b$ are the endpoints of an external path in $G$ but
$a'$ and $b'$ in $G'$ are not. Then Spoiler wins with 9 pebbles in $2\log n+2$ moves.
The same holds true with the roles of $G$ and $G'$ interchanged.
\end{lemma}

\begin{proof}
Suppose that both $a'$ and $b'$ are in $H'$ because otherwise Spoiler wins
by Lemma \ref{lem:srespect}.
Using the threat given by Lemma \ref{lem:srespect}, Spoiler restricts play
to $G\setminus(V(H)\setminus\{a,b\})$ and $G'\setminus(V(H')\setminus\{a',b'\})$.
In these graphs we have $d(a,b)<n$ while $d(a',b')=\infty$ and Spoiler wins
by following the halving strategy of Lemma~\ref{lem:halving}.
\end{proof}

For the notion of an external path, we now introduce a geometrical (rather than a logical) 
specification.
The boundaries of $\calS_0(x,u)$ and $\calS_0(y,v)$ compose to give
the topological boundary of $H$ (which does not depend on a particular embedding). 
Its segment between $x$ and $u$ will be called
the $x$-side of $H$ and the segment between $y$ and $v$ will be called
the $y$-side. To be precise, we define \emph{the $x$-side of $H$} to be the path
$B_1(x,u)$ and \emph{the $y$-side of $H$} to be the path $B_1(y,v)$. In $G'$,
the $x$-side of $H'$ will refer to the boundary segment between $x'$ and $v'$
and the $y$-side of $H'$ will refer to the segment between $y'$ and $u'$.
Formally, \emph{the $x$-side of $H'$} is defined to be the path
$B_1(x',u')[x',v'_1]B_2(y',v')[v'_1,v']$ and \emph{the $y$-side of $H'$} to be the path 
$B_1(y',v')[y',u'_1]B_2(x',u')[u'_1,u']$, see Figure \ref{fig:4}. 
Clearly, an external path can connect only
vertices on the same side of $H$ or $H'$. We will correspondingly distinguish 
\emph{$x$-external} paths and \emph{$y$-external} paths.

The strategy will be based on a sequence of vertices $c_1,a_1,b_1,\ldots,c_m,a_m,b_m$
and a sequence of external paths $P_1,\ldots,P_m$, where $m\le l\le d_0(z_1,w_1)+1$, in $G$
and similar sequences in $G'$. While the $c_i,a_i,b_i$'s will be defined uniquely, for the $P_i$'s
we will fix one of possibly several choices. The $c_1,a_1,b_1,P_1$ are defined by the
following conditions. The first four are to be fulfilled unconditionally.
The other are to be fulfilled in the given order.

\begin{enumerate}
\item
$c_1=e_1$.
\item
$d_0(x,a_1)<d_0(x,c_1)$.
\item
$d_0(x,b_1)>d_0(x,c_1)$.
\item
$P_1$ is an external $a_1$-$b_1$-path. (Such a path exists by the assumption that
$G$ is triconnected because otherwise $G\setminus\{z,c_1\}$ would be disconnected).
\item
If possible, $P_1$ is $x$-external.
\item
$d_0(x,a_1)$ is maximum possible. ($a_1$ is therewith uniquely determined.)
\comm{`minimum' instead of `maximum' would be equally good.}
\item
$d_0(x,b_1)$ is maximum possible. ($b_1$ is therewith uniquely determined.)
\comm{`maximum' is essential for the proof.}
\end{enumerate}

\begin{lemma}{\assfont[Assumptions \ref{ass:d0equal}, \ref{ass:cuts}, \ref{ass:essential}]}%
\label{lem:abc1respect}
\begin{bfenumerate}
\item
If Spoiler pebbles $c_1$, Duplicator is forced to respond with $c'_1$ because
otherwise Spoiler wins with 9 pebbles in extra $\log n+2$ moves.
\item
If $P_1$ is $x$-external but $P'_1$ is $y$-external or vice versa,
Spoiler wins with 13 pebbles in $3\log n+11$ moves.
\item
Suppose that $P_1$ and $P'_1$ are simultaneously $x$- or $y$-external.
If Spoiler pebbles $a_1$ and $b_1$, Duplicator responds with, respectively, $a'_1$
and $b'_1$ or Spoiler wins with 13 pebbles in extra $3\log n+12$ moves.
(In the last sentence, `or' is not exclusive, say,
Spoiler wins if $d_0(x',a'_1)\ne d_0(x,a_1)$ or $d_0(x',b'_1)\ne d_0(x,b_1)$.)
\end{bfenumerate}
\end{lemma}

\begin{proof}
{\bf 1.}
The assumptions ensure that, if Duplicator responds not with $c'_1$, she violates
the $d_0$-distance and hence Spoiler wins by Lemma~\ref{lem:d0respect}.

{\bf 2.}
To be specific, suppose that $P_1$ is $x$-external while $P'_1$ is $y$-external
(the other case is symmetric). 
Spoiler pebbles $a_1$, $b_1$, and $c_1$.
Denote Duplicator's responses by $a'$, $b'$, and $c'$.
By Item 1, we assume $c'=c'_1$. If $a'$ or $b'$ is outside $H'$,
Spoiler has a fast win by Lemma \ref{lem:srespect}, so let $a',b'\in H'$.
To make Spoiler's life harder, assume that the $d_0$-distances between
the pebbled vertices in $G$ and $G'$ agree. It follows that
$a'$ is in $H'_0$ but $b'$ is not.
If there is no external $a'$-$b'$-path, Spoiler wins by Lemma \ref{lem:extrespect}.
Suppose there is one. By assumption, $a'$ lies on the $y$-side of $H'_0$
and Lemma \ref{lem:siderespect} applies.

{\bf 3.}
To be definite, suppose that $P_1$ and $P'_1$ are $x$-external.
Let Spoiler pebble $a_1$ and $b_1$. As was shown in the proof of Item 2,
Duplicator is forced to respond with $a'\in H'_0$ and $b'\notin H'_0$ being the endpoints
of an $x$-external path. If $a'\ne a'_1$, Spoiler pebbles $a'_1$ and $b'_1$.
By the definition of $a'_1$, we have $d_0(x',a'_1)>d_0(x',a')$.
Duplicator should now respond with $a^*\in H_0$ with $d_0(x,a^*)>d_0(x,a_1)$ and 
$b^*\notin H_0$ being the endpoints of an $x$-external path. It is not hard to see
that this is possible only if the $d_0$ distance is violated and hence Spoiler
wins fast.
If $a'=a'_1$ but $b'\ne b'_1$, Spoiler wins following a similar strategy.
\end{proof}

Suppose that $c_i,a_i,b_i$ are already defined and $P_i$ is already fixed.
If $b_i$ is in $H_l$ and $b_i\ne e_l$, set $m=i$ and terminate.
Otherwise define $c_{i+1},a_{i+1},b_{i+1}$ and $P_{i+1}$ as follows.

\begin{enumerate}
\item
If $b_i$ is an essential cutpoint, then $c_{i+1}=b_i$.
If $b_i$ is not an essential cutpoint and $b_i\in H_j$, then $c_{i+1}$
is the essential cutpoint in $H_j$ nearer to $w_1$
(i.e., $d_0(x,c_{i+1})>d_0(x,b_i)$).
\item
$d_0(x,a_{i+1})<d_0(x,c_{i+1})$.
\item
$d_0(x,b_{i+1})>d_0(x,c_{i+1})$.
\item
$P_{i+1}$ is an external $a_{i+1}$-$b_{i+1}$-path. (Such a path exists by the assumption that
$G$ is triconnected because otherwise $G\setminus\{z,c_{i+1}\}$ would be disconnected).
\item
If $P_i$ is $x$-external (resp.\ $y$-external), then $P_{i+1}$ is $x$-external 
(resp.\ $y$-external) too, provided such a choice exists.
\item
$d_0(x,a_{i+1})$ is maximum possible. ($a_{i+1}$ is therewith uniquely determined.)
\item
$d_0(x,b_{i+1})$ is maximum possible. ($b_{i+1}$ is therewith uniquely determined.)
\end{enumerate}

\begin{lemma}{\assfont[Assumptions \ref{ass:d0equal}, \ref{ass:cuts}, \ref{ass:essential}]}%
\label{lem:abirespect}
Let $i<m$. Suppose that $a_i$ and $b_i$ are pebbled in $G$ while
$a'_i$ and $b'_i$ are pebbled correspondingly in $G'$
and that $P_i$ and $P'_i$ are both $x$-external or both $y$-external.
\begin{bfenumerate}
\item
Suppose that Spoiler pebbles $c_{i+1}$, $a_{i+1}$, and $b_{i+1}$.
Then Duplicator responds, respectively, with $c'_{i+1}$, $a'_{i+1}$, and $b'_{i+1}$
or Spoiler wins with 13 pebbles in extra $3\log n+14$ moves.
\item
In the last sentence, `or' is not exclusive. In particular,
Spoiler wins if $P_{i+1}$ and $P'_{i+1}$ are not both $x$-external or not both $y$-external.
\item
Similarly to Item 1, Duplicator is forced to respond with $c_{i+1}$, $a_{i+1}$, and $b_{i+1}$
if Spoiler pebbles $c'_{i+1}$, $a'_{i+1}$, and $b'_{i+1}$.
\comm{Is Item 3 a logical consequence of Item 1?}
\end{bfenumerate}
\end{lemma}

\begin{proof}
Denote Duplicator's responses by $c'$, $a'$, and $b'$.
We assume that $d_0(x,a_i)=d_0(x',a'_i)$ and $d_0(x,b_i)=d_0(x',b'_i)$
for else Spoiler wins by Lemma \ref{lem:d0respect}.
It follows by Assumption \ref{ass:cuts} that $c'_{i+1}$ is uniquely determined
by the conditions $d_0(x',c'_{i+1})=d_0(x,c_{i+1})$ and $d_0(u',c'_{i+1})=d_0(u,c_{i+1})$. 
Thus, if $c'\ne c'_{i+1}$, the distance $d_0$ gets violated and Duplicator loses soon.
Assume hence that $c'=c'_{i+1}$. Now Spoiler checks if $a'=a'_{i+1}$ and $b'=b'_{i+1}$
by tracing through Conditions 2--7 of the definition above and uses any disagreement
to win. To exclude one of Spoiler's threats, we will assume that $d_0(x,a')=d_0(x,a_{i+1})$ 
and $d_0(x,b')=d_0(x,b_{i+1})$.

Conditions 2 or 3 are true because otherwise the $d_0$ distance would be violated.
If Condition 4 is violated, Spoiler wins by Lemma \ref{lem:extrespect}.
Let $P'$ be an external $a'$-$b'$-path.
Verification of Condition 5 splits into two cases.
The same argument will prove Item 2 of the lemma.
For definiteness, we assume without loss of generality that $P_i$ is $x$-external,
see Figure~\ref{fig:10}.

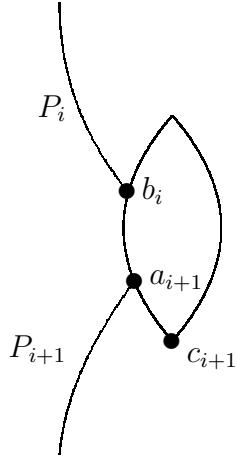
\begin{figure}[htbp]
\centerline{
\unitlength=1.00mm
\special{em:linewidth 0.4pt}
\linethickness{0.4pt}
\begin{picture}(43.00,65.00)
\bezier{450}(30.00,50.00)(17.00,35.00)(30.00,20.00)
\bezier{450}(30.00,50.00)(43.00,35.00)(30.00,20.00)
\bezier{112}(24.00,40.00)(15.00,51.00)(15.00,65.00)
\put(24.00,40.00){\circle*{2.00}}
\put(30.00,20.00){\circle*{2.00}}
\bezier{104}(25.00,28.00)(16.00,16.00)(15.00,5.00)
\put(25.00,28.00){\circle*{2.00}}
\put(26.00,40.00){\makebox(0,0)[lc]{$b_i$}}
\put(27.00,28.00){\makebox(0,0)[lc]{$a_{i+1}$}}
\put(32.00,19.00){\makebox(0,0)[lt]{$c_{i+1}$}}
\put(16.00,17.00){\makebox(0,0)[rb]{$P_{i+1}$}}
\put(16.00,51.00){\makebox(0,0)[rc]{$P_i$}}
\end{picture}
}
\caption{Proof of Lemma \protect\ref{lem:abirespect} (Subcase 1.2).}%
\label{fig:10}
\end{figure}

\case 1{$c_{i+1}\ne b_i$.}
Let $H_j$ be the block of $H$, containing $b_i$ and $c_{i+1}$ 
(as in Condition 1 defining $c_{i+1}$).

\subcase{1.1}{$P_{i+1}$ is $x$-external.}
Note that $a_{i+1}$ is on the boundary of $H_j$ between $b_i$ and $c_{i+1}$
(the equality $a_{i+1}=b_i$ is possible).
The $a_{i+1}$ cannot precede $b_i$, in particular, belong to any other block,
because in this case $P_i$ could be prolonged (contradictory to Condition 7 defining $b_i$).
Since $a_{i+1}\in H_j$, by Assumption \ref{ass:cuts} we have $a'\in H'_j$.
If $a'$ and $b'_i$ are on the opposite boundaries of $H'_j$,
Spoiler wins by the Basic Stratagem of Lemma \ref{lem:stratagem}.
Otherwise Condition 5 is met.

\subcase{1.2}{$P_{i+1}$ is $y$-external.}
If $P'_{i+1}$ is $x$-external, Spoiler pebbles $a'_{i+1}$ and $b'_{i+1}$ 
and wins quite similarly to Subcase 1.1 using
the Basic Stratagem. Otherwise $P'$ is $y$-external and Condition 5 is met.

\case 2{$c_{i+1}=b_i$.}
Suppose that $c'_{i+1}=b'_i$ for else Duplicator has already lost.
Condition 5 is met because neither $P_{i+1}$ nor $P'$ can be $x$-external
(otherwise $P_i$ or $P'_i$ could be prolonged).

Finally, if Condition 6 or 7 is false, Spoiler wins 
similarly to the proof of Lemma~\ref{lem:abc1respect}.3.
\end{proof}

Finally, we are prepared to describe Spoiler's strategy.
In the first two rounds Spoiler pebbles $a_1$, $b_1$, and $c_1$.
In view of Lemma \ref{lem:abc1respect}, we let Duplicator respond with
$a'_1$, $b'_1$, and $c'_1$ and assume that $d_0(x',a'_1)=d_0(x,a_1)$, 
$d_0(x',b'_1)=d_0(x,b_1)$, and $d_0(x',c'_1)=d_0(x,c_1)$.

Suppose first that $d_0(x',a'_i)\ne d_0(x,a_i)$, $d_0(x',b'_i)\ne d_0(x,b_i)$,
or $d_0(x',c'_i)\ne d_0(x,c_i)$
for some $i\ge2$. Let $\ell$ be the smallest index with this property.
Spoiler pebbles $a_\ell$, $b_\ell$, and $c_\ell$. If Duplicator responds with 
$a'_\ell$, $b'_\ell$, and $c'_\ell$, she violates the $d_0$-distance and 
loses by Lemma \ref{lem:d0respect}. Otherwise Spoiler follows the Generalized
Halving Strategy of Lemma \ref{lem:exthalving}.2 and,  
by Lemmas \ref{lem:exthalving} and \ref{lem:abirespect},
wins with 15 pebbles making less than $6+3(\log\ell+1)+3\log n+14\le6\log n+23$ 
moves at total.

Suppose therefore that $d_0(x',a'_i)=d_0(x,a_i)$, $d_0(x',b'_i)=d_0(x,b_i)$,
and $d_0(x',c'_i)=d_0(x,c_i)$
for all $i\le m$. Call an index $i$ \emph{distinguishing} if $P_i$ is $x$-external
but $P'_i$ is $y$-external or vice versa. Let $\ell$ be the smallest distinguishing index.
If $\ell=1$, Spoiler wins by Lemma \ref{lem:siderespect}. If $\ell\ge2$, Spoiler selects
$a_{\ell-1}$, $b_{\ell-1}$, $c_{\ell-1}$, $a_{\ell}$, $b_{\ell}$, and $c_{\ell}$.
If Duplicator responds with 
$a'_{\ell-1}$, $b'_{\ell-1}$, $c'_{\ell-1}$, $a'_{\ell}$, $b'_{\ell}$, and $c'_{\ell}$,
she loses by Lemma \ref{lem:abirespect}.2.
Otherwise Spoiler uses the Generalized Halving Strategy
and, by Lemma \ref{lem:exthalving}.2, wins with 15 pebbles making less than 
$9+3(\log\ell+1)+3\log n+14\le6\log n+26$ moves at total.

There remains the case that no $i$ is distinguishing. Without loss of generality,
assume that both $P_m$ and $P'_m$ are $x$-external.
This means that $b_m$ lies on $B_1(e_m,u)$ and $b'_m$ lies on $B_2(e'_m,v')$.
Spoiler pebbles $a_m$, $b_m$, and $c_m$. If Duplicator responds with
$a'_m$, $b'_m$, and $c'_m$, she loses by Lemma \ref{lem:siderespect}.
Otherwise Spoiler wins by using the Generalized Halving Strategy.
Lemma \ref{lem:colltwist} is proved.

\section{Global strategy}\label{s:global}

This section contains the second part of the proof of Theorem \ref{thm:dist3},
which was started in Section \ref{s:local}.
Let $G$ and $G'$ be non-isomorphic triconnected planar graphs.
We have to design a strategy allowing Spoiler to win the \EF\/ game
on $G$ and $G'$ with 15 pebbles in less than $11\log n+43$ rounds.
By the Whitney theorem, every triconnected planar graph has a unique
embedding in the sphere. We use two combinatorial specifications for
the concept of an embedding. One is a standard notion of a \emph{rotation system}
(see, e.g.\ \cite{MTh}). The other is a related, but in a sense ``poorer'',
notion of a \emph{layout system} (see Subsections \ref{ss:RLdef} 
and \ref{ss:RLGdef} for the definitions).
Denote the rotation and the layout systems for $G$ and $G'$ by $R$ and $R'$
and by $L$ and $L'$ respectively. In Subsection \ref{ss:rotation} we will
show that every rotation system is succinctly definable, in particular,
Spoiler has an efficient winning strategy in the \EF\/ game on $R$ and $R'$.
In Subsection \ref{ss:LtoR} we will see that Spoiler can win the game on $L$ and $L'$
by emulating the game on $R$ and $R'$. In its turn, our main achievement
of Section \ref{s:local}, Lemma \ref{lem:colltwist}, allows Spoiler to win
the game on $G$ and $G'$ by emulating the game on $L$ and $L'$. This emulation
is presented in Subsection \ref{ss:GtoL}. With these preliminaries, the proof
of Theorem \ref{thm:dist3} in Subsection \ref{ss:proofdist3} takes no efforts.

\subsection{Two specifications of a graph embedding}\label{ss:RLdef}

The following definitions are introduced for a connected graph $G$ with
minimum vertex degree at least~3.

A \emph{rotation system} $R=\langle G,T\rangle$ is a structure 
consisting of a graph $G$ and a ternary relation $T$ on $V(G)$
satisfying the following conditions:

\begin{enumerate}
\item
If $T(a,b,c)$, then $b$ and $c$ are in $\Gamma(a)$, the neighborhood of $a$ in $G$.
\item
For every $a$ the binary relation $T_a(b,c)=T(a,b,c)$ is a directed cycle
on $\Gamma(a)$ (i.e., for every $b$ there is exactly one $c$ such that $T_a(b,c)$
for every $c$ there is exactly one $b$ such that $T_a(b,c)$,
and the digraph $T_a$ is connected).
\end{enumerate}

Geometrically, $T_a$ describes
the circular order in which the edges of $G$ incident to $a$ occur in the embedding
if we go around $a$ clockwise.

Given a rotation system $R=\langle G,T\rangle$, we define another rotation system
$R^*=\langle G,T^*\rangle$ by $T^*_a(b,c)=T_a(c,b)$ and call it the \emph{conjugate}
of $R$. Geometrically, $R^*$ is a variant of $R$ if we look at $R$ from the other
side of the surface. Obviously, $(R^*)^*=R$.

A \emph{layout system} $L=\langle G,T,Q\rangle$ is a structure 
consisting of a graph $G$ and two relations on $V(G)$, ternary $T$ and quaternary $Q$,
satisfying the following conditions.
\begin{enumerate}
\item
If $T(a,b,c)$, then $b$ and $c$ are in $\Gamma(a)$, the neighborhood of $a$ in $G$.
Furthermore, for every $a$ the binary relation $T_a(b,c)=T(a,b,c)$ is an undirected 
cycle on $\Gamma(a)$ (that is, $T_a$ is symmetric, irreflexive, and connected).
\item
If $Q(b_1,a_1,a_2,b_2)$, then $b_1,a_1,a_2,b_2$ is a path in $G$ or, if $b_1=b_2$, 
it is a cycle.
Every pair $(a_1,a_2)$ with $a_1$ and $a_2$ adjacent in $G$ extends to exactly two
quadruples $(b_1,a_1,a_2,b_2)$ and $(c_1,a_1,a_2,c_2)$ satisfying $Q$.
Moreover, for both $i=1,2$, the $b_i$ and $c_i$ are the neighbors of $a_{3-i}$ 
in the cycle $T_{a_i}$,
that is, $T(a_i,a_{3-i},b_i)$ and $T(a_i,a_{3-i},c_i)$ are both true.
\end{enumerate}

Relations $T$ and $Q$ also have clear geometric meaning. Namely, $T_a$ determines
the (undirected) circular order in which the edges of $G$ incident to $a$ are embedded.
Note that now we specify no clockwise (or counter-clockwise) direction around $a$.
This is the point where a layout system deviates from
a rotation systems. Thus, if a vertex $a_1$ and its neighborhood
are already embedded and $a_2$ is adjacent to $a_1$, we have still two different ways
to embed the neighborhood of $a_2$. The proper choice is determined by $Q$.
Namely, it is supposed that the facial cycle going via $b_1,a_1,a_2$
goes further via $b_2$ and the facial cycle going via $c_1,a_1,a_2$
goes further via $c_2$.

Given a rotation system $R=\langle G,T_R\rangle$, we associate with it
a layout $\lay(R)=\langle G,T_L,Q\rangle$ according to the geometric meaning.
Namely, $T_L$ is defined by $T_L(a,b,c)=T_R(a,b,c)\vee T_R(a,c,b)$.
To define $Q$, we first introduce the successor and the predecessor functions
on $\Gamma(a)$, $s_a$ and $p_a$, by the equalities $c=s_a(b)$ and $b=s_a(c)$ 
if $T_R(a,b,c)=1$. Now we set the following two relations true:
$Q(p_{a_1}(a_2),a_1,a_2,s_{a_2}(a_1))$ and $Q(s_{a_1}(a_2),a_1,a_2,p_{a_2}(a_1))$.
As easily seen, $\lay(R)=\lay(R^*)$.

Let $L=\lay(R)$. The following lemma shows that the pair $\{R,R^*\}$
is reconstructible from $L$.

\begin{lemma}\label{lem:laylay}
If $\lay(R')=\lay(R)$, then either $R'=R$ or $R'=R^*$.
\end{lemma}

\begin{proof}
Suppose, to the contrary, that neither $R'=R$ nor $R=R^*$.
Since $G$ is connected, there are adjacent vertices $a_1$ and $a_2$
such that the relation $T'_{a_1}$ is different from $T_{a_1}$
and $T'_{a_2}$ is different from $T^*_{a_2}$, which means that
$T'_{a_1}$ is identical to $T^*_{a_1}$ and $T'_{a_2}$ is identical to $T_{a_2}$.
From here it is easy to infer that, if we denote the quaternary relations in 
$\lay(R')$ and $\lay(R)$ by $Q$ and $Q'$ respectively, then the binary relations
$Q(\cdot,a_1,a_2,\cdot)$ and $Q'(\cdot,a_1,a_2,\cdot)$ are not identical,
a contradiction.
\end{proof}

In fact, Lemma \ref{lem:b} is essentially strengthened below, see
Lemma \ref{lem:laylay}.

\subsection{Defining a rotation system}\label{ss:rotation}

The material of this subsection is borrowed from \cite{GVe}.

\begin{theorem}\label{lem:rot}
For every rotation system $R=\langle G,T\rangle$, we have $D^{5}(R)<3\log n+8$.
\end{theorem}

The proof takes the rest of this subsection.
The proof is based on Equality \refeq{eq:dddc} and Proposition \ref{prop:game}.
Let $R=\langle G,T\rangle$ be a rotation system with $n$ vertices
and $R'=\langle G',T'\rangle$ be a non-isomorphic structure of the same signature.
We have to design a strategy for Spoiler
in the  Ehrenfeucht-Fra\"\i{}ss\'{e} game on $R$ and $R'$ allowing him to win with only
5 pebbles in less than $3\log n+8$ moves, whatever Duplicator's strategy.

The case that $R'$ is not a rotation system is simple. Spoiler needs just 4 moves
to show that $R'$, unlike $R$, does not fit the definition
(which has quantifier depth 4).
We will therefore suppose that $R'$ is a rotation system as well.

The main idea of the proof is to show that a rotation system admits a natural coordinatization
and that Duplicator must respect vertex coordinates. A coordinate system on $R=\langle G,T\rangle$
is determined by fixing its origins, namely, an ordered edge of $G$.
We first define \emph{local coordinates} on the neighborhood of a vertex $x$.
Fix $y\in\Gamma(x)$ and let $z$ be any vertex in $\Gamma(x)$. Then $c_{xy}(z)$
is defined to be the number of $z$ in the order of $T_x$ if we start counting from $c_{xy}(y)=0$.
In the global system of coordinates specified by an ordered pair of adjacent $a,b\in V(G)$,
each vertex $v\in V(G)$ receives coordinates $C_{ab}(v)$ defined as follows.
Given a path $P=a_0a_1a_2\ldots a_l$ from $a_0=a$ to $a_l=v$,
let $C_{ab}(v;P)=(c_1,\ldots,c_l)$ be a sequence of integers with 
$c_1=c_{ab}(a_1)$ and $c_i=c_{a_{i-1}a_{i-2}}(a_i)$ for $i\ge2$.
We define $C_{ab}(v)$ to be the lexicographically minimum $C_{ab}(v;P)$ over all $P$.
Note that $C_{ab}(v)$ has length $d(a,v)$.
By $P_v$ we will denote the path for which $C_{ab}(v)=C_{ab}(v;P_v)$.
One can say that $P_v$ is \emph{the extreme left shortest path from $a$ to $v$}.
Note that $P_v$ is reconstructible from $C_{ab}(v)$ and hence 
different vertices receive different coordinates.
The following observation enables a kind of the halving strategy.

\begin{lemma}\label{lem:split}
Let $a,b,v\in V(G)$ and $a',b',v'\in V(G')$, where $a$ and $b$ as well as
$a'$ and $b'$ are adjacent.
Assume that $d(a,v)=d(a',v')$ but $C_{ab}(v)\ne C_{a'b'}(v')$.
Furthermore, let $u$ and $u'$ lie on $P_v$ and $P_{v'}$ at the same distance
from $a$ and $a'$ respectively. Assume that $C_{ab}(u)=C_{a'b'}(u')$. 
Finally, let $w$ and $w'$ be predecessors of
$u$ and $u'$ on $P_v$ and $P_{v'}$ respectively.
Then $C_{uw}(v)\ne C_{u'w'}(v')$.
\end{lemma}

\begin{proof}
By definition, $C_{ab}(v)=C_{ab}(u)C_{uw}(v)$ and $C_{a'b'}(v')=C_{a'b'}(u')C_{u'w'}(v')$.
\end{proof}

\begin{lemma}\label{lem:coord}
Suppose that $a,b,v\in V(G)$ and $a',b',v'\in V(G')$
are pebbled coherently to the notation.
Assume that $a$ and $b$ as well as $a'$ and $b'$ are adjacent
and that $C_{ab}(v)\ne C_{a'b'}(v')$.
Then Spoiler is able to win with 5 pebbles in less than $3\log n+3$ moves.
\end{lemma}

\begin{proof}
Assume that $d(a,v)\ge2$.
If $d(a,v)\ne d(a',v')$, Spoiler wins in less than $\log n+1$ moves
by Lemma \ref{lem:halving}. If $d(a,v)=d(a',v')$, Spoiler applies a more
elaborated halving strategy.
Let $u$ be the vertex on $P_v$ with
$
d(a,u)=\lceil d(a,v)/2\rceil
$
and $u'$ be the corresponding vertex on $P_{v'}$.

\case 1{$C_{ab}(u)\ne C_{a'b'}(u')$.}
Without loss of generality assume that $C_{ab}(u)$ is lexicographically smaller
than $C_{a'b'}(u')$ (otherwise Spoiler moves in the other graph symmetrically).
Spoiler pebbles $u$. Denote Duplicator's response in $G'$ by $u^*$.
If $C_{ab}(u)=C_{a'b'}(u^*)$, then in our coordinate system $u^*$
is strictly on the left side to $P_{v'}$, the ``left most'' shortest path
from $a'$ to $v'$. It follows that $d(u^*,v')>d(u',v')=d(u,v)$ and
Spoiler wins fast by Lemma \ref{lem:halving}.
If $C_{ab}(u)\ne C_{a'b'}(u^*)$, then Spoiler has the same configuration as
at the beginning, with $u,u^*$ in place of $v,v'$, and with the distance
$d(a,u)$ twice reduced if compared to $d(a,v)$.
Then Spoiler does all the same once again.

\case 2{$C_{ab}(u)=C_{a'b'}(u')$.}
Spoiler pebbles $u$. If Duplicator responds with $u^*\ne u'$ then either
$d(a,u)\ne d(a',u)$ or $d(a,u)=d(a',u)$ but $C_{ab}(u)\ne C_{a'b'}(u^*)$
and Spoiler has a configuration similar to the beginning.
Assume therefore that $u^*=u'$.

Let $w$ and $w'$ be as in Lemma \ref{lem:split}.
Now Spoiler acts with $w,w'$ exactly in the same way as he just did with $u,u'$.
As a result, the players pebble vertices $\tilde w\in V(G)$ and
$\tilde w'\in V(G')$, where $\tilde w=w$ or $\tilde w'=w'$, with three possible outcomes:
\begin{enumerate}
\item
Some distances between the corresponding vertices in $G$ and $G'$ disagree.
\item
Spoiler achieves the same configuration as at the beginning with $\tilde w,\tilde w'$
in place of $v,v'$, where $d(a,\tilde w)<\lceil d(a,v)/2\rceil$.
\item
$\tilde w=w$ and $\tilde w'=w'$.
\end{enumerate}
In the first case Spoiler wins fast. In the third case Lemma \ref{lem:split}
applies and again Spoiler has the same configuration as at the beginning
with respect to new coordinate origins $(u,w)$ and $(u',w')$,
where $d(u,v)=\lfloor d(a,v)/2\rfloor$ is reduced.

In less than $\log d(a,v)+1$ iterations Spoiler forces a configuration as at the beginning
with $d(a,v)=1$ (we restore the initial notation), so it remains to consider this case.
Suppose that $d(a',v')=1$ as well. Now we have disagreement of local coordinates:
$c_{ab}(v)\ne c_{a'b'}(v')$. Keeping the pebbles on $a$ and $a'$, Spoiler restricts
play to the directed cycles $T_a$ and $T'_{a'}$ and wins with other 3 pebbles 
in less than $\log\deg a+1$
moves applying an analog of the strategy of Lemma \ref{lem:halving} for linear orders.

Each iteration takes at most 2 moves, which may be needed in Case 2.
Thus, Spoiler needs less than $2(\log\diam G+1)+(\log\Delta(G)+1)\le3\log n +3$
moves to win. The maximum number of pebbles is on the board in Case 2
(on $a,b,v,u,$ and $w$).
\end{proof}

Now we are ready to describe Spoiler's strategy in the game.
In the first two rounds he pebbles $a$ and $b$, arbitrary adjacent vertices in $G$.
Let Duplicator respond with adjacent $a'$ and $b'$ in $G'$.
If $G$ contains a vertex $v$ with coordinates $C_{ab}(v)$ different from
every $C_{a'b'}(v')$ in $G'$ or if $G'$ contains a vertex with
coordinates absent in $G$, then Spoiler pebbles it and wins by Lemma \ref{lem:coord}.
Suppose therefore that the coordinatization determines a matching between $V(G)$ and $V(G')$.
Given $x\in V(G)$, let $f(x)$ denote the vertex $x'\in V(G')$ with $C_{a'b'}(x')=C_{ab}(x)$.
If $f$ is not an isomorphism from $G$ to $G'$, then Spoiler pebbles two vertices
$u,v\in V(G)$ such that the pairs $u,v$ and $f(u),f(v)$ have different adjacency.
Not to lose immediately, Duplicator responds with a vertex having different coordinates
and again Lemma \ref{lem:coord} applies.
If $f$ is an isomorphism between $G$ and $G'$, then this map does not respect the
relations $T$ and $T'$ and Spoiler demonstrates this similarly.
The proof of Theorem \ref{lem:rot} is complete.

\subsection{Reducing the play on layout systems to the play on rotation systems}\label{ss:LtoR}

We start with an auxiliary lemma.

\begin{lemma}\label{lem:a}
Let $R=\langle G,T'\rangle$ and $R'=\langle G,T\rangle$ be rotation systems. 
Let $L=\lay(R)$ and $L'=\lay(R')$. Suppose that, while 
$T(a_1,b_1,c_1)=T(a_2,b_2,c_2)=1$ in $R$, in $R'$ we have
$T'(a'_1,b'_1,c'_1)=T'(a'_2,c'_2,b'_2)=1$. 
Then Spoiler wins 
$\game^9_{2\log n+4}(L,a_1,b_1,c_1,a_2,b_2,c_2,L',a'_1,b'_1,c'_1,a'_2,b'_2,c'_2)$.
\end{lemma}

\begin{proof}
\case 1{$a_1=a_2=a$.}
Correspondingly, suppose that $a'_1=a'_2=a'$.
The case that $\{b_1,c_1\}$ and $\{b_2,c_2\}$ intersect is simple;
we hence suppose that all these vertices are pairwise distinct.
Spoiler restricts play to the graphs $T_a\setminus\{b_1,b_2\}$
and $T'_a\setminus\{b'_1,b'_2\}$ (those are actually directed graphs
but, if Spoiler ignores the edge directions and wins the game on the
corresponding graphs, this implies his win on the digraphs).
In these graphs $d(c_1,c_2)=\infty$ while $d(c'_1,c'_2)<\infty$
and hence Spoiler wins in less than $\log\deg a+1$ moves.

\case 2{$a_1$ and $a_2$ are adjacent.}
It suffices to consider a special subcase where
$b_1=a_2$ and $b_2=a_1$. Spoiler can force either this subcase or Case 1 
in 2 extra moves. By the definition of $\lay(R)$, we have $Q(c_1,a_1,a_2,c_2)=0$
whereas $Q'(c'_1,a'_1,a'_2,c'_2)=1$, which is a win for Spoiler.

\case 3{$d(a_1,a_2)\ge2$.}
Spoiler reduces this case to Case 2 in $\lceil\log d(a_1,a_2)\rceil$ moves.
He first pebbles a vertex $a_3$ on the midway between $a_1$ and $a_2$
and then two more vertices $b_3,c_3$ so that $T(a_3,b_3,c_3)=T(a_i,b_i,c_i)$, $i=1,2$.
For Duplicator's response $a'_3,b'_3,c'_3$, assume that one of the relations
$T'(a'_3,b'_3,c'_3)$ or $T'(a'_3,c'_3,b'_3)$ is true for else Spoiler has already won.
We have either
$T'(a'_3,b'_3,c'_3)\ne T'(a'_1,b'_1,c'_1)$ or $T'(a'_3,b'_3,c'_3)\ne T'(a'_2,b'_2,c'_2)$.
In either case, one of the tuples $(a_i,b_i,c_i,a_3,b_3,c_3)$, for $i=1$ or $i=2$,
is similar to the initial position, while the distance between the two $a$-vertices
has decreased. Spoiler just iterates this tricks sufficiently many times.
\end{proof}

\begin{lemma}\label{lem:b}
Let $R=\langle G,T'\rangle$ and $R'=\langle G,T\rangle$ 
be rotation systems such that neither $R'\cong R$ nor $R'\cong R^*$. 
Suppose that $m\ge\max\{W(R,R'),W(R^*,R')\}$ and set $k=3+\max\{m,6\}$.
Let $L=\lay(R)$ and $L'=\lay(R')$. Then 
$$
D^k(L,L')<\max\{D^m(R,R'),D^m(R^*,R')\}+2\log n+7.
$$
\end{lemma}

\begin{proof}
We design a strategy for Spoiler in $\game^k(L,L')$.
In the first three rounds he pebbles vertices $a_0,b_0,c_0$ in $V(G)$ so that
$T(a_0,b_0,c_0)=1$. Denote Duplicator's responses by $a'_0,b'_0,c'_0$
and suppose that either $T'(a'_0,b'_0,c'_0)=1$ or $T'(a'_0,c'_0,b'_0)=1$
(otherwise Spoiler has won). Without loss of generality, suppose the former
(otherwise just interchange $b_0$ and $c_0$ and consider $R^*$ and $T^*$ instead
of $R$ and $T$). Starting from the 4-th round, Spoiler emulates $\game^m(R,R')$.
His win in this game means that either the equality, or the adjacency in $G$,
or the ternary relation is violated. The former two cases imply also Spoiler's
win in $\game^k(L,L')$. In the latter case we arrive at the conditions of
Lemma \ref{lem:a} and Spoiler needs no more than $2\log n+4$ moves to win. 
\end{proof}

\subsection{The layout and the rotation system of a triconnected planar graph}\label{ss:RLGdef}

Let $\sigma$ be an embedding of a graph $G$ in a sphere.
Recall that, by definition, $\sigma$ is an isomorphism from $G$
to a sphere graph $\tilde G$. We define the rotation system 
$R_\sigma=\langle G,T_\sigma\rangle$ according to a natural geometric meaning. 
Namely, for $a\in V(G)$ and $b,c\in\Gamma(a)$ we have $T_\sigma(a,b,c)=1$
if, looking at the neighborhood of $\sigma(a)$ in $\tilde G$ from the standpoint
at the sphere center, $\sigma(b)$ is followed by $\sigma(c)$ in the clockwise order.
Note that $R^*_\sigma$ corresponds to the view on $\tilde G$ from the outside.
We can define the layout system $L_\sigma$ also geometrically, as described in
Subsection \ref{ss:RLdef}. Equivalently, we set $L_\sigma=\lay(R_\sigma)$.

Let $\sigma\function G{\tilde G}$ and $\tau\function G{\hat G}$
be two spherical embedding of $G$. Suppose that they are equivalent,
that is, $\tau\circ\sigma^{-1}$ is induced by a homeomorphism from the sphere
where $\tilde G$ is drawn to the sphere where $\hat G$ is drawn.
Since $\tau\circ\sigma^{-1}$ takes a facial cycle to a facial cycle,
we have $L_\sigma=L_\tau$. By Lemma \ref{lem:laylay}, we also have
$\{R_\sigma,R^*_\sigma\}=\{R_\tau,R^*_\tau\}$.

Given a triconnected planar graph $G$, we define $L_G=L_\sigma$
and $R_G=R_\sigma$ for $\sigma$ being an arbitrary embedding of $G$
in a sphere. By the Whitney theorem, the definition does not depend on
a particular choice of $\sigma$ 
if we agree that $R_G$ is defined up to taking the conjugate.

\subsection{Reducing the play on graphs to the play on layout systems}\label{ss:GtoL}

\begin{lemma}\label{lem:aa}
Suppose that $G$ and $G'$ are non-isomorphic triconnected planar graphs.
Let $L_G=\langle G,T,Q\rangle$ and $L_{G'}=\langle G',T',Q'\rangle$.
\begin{bfenumerate}
\item
If $T(a,b,c)\ne T'(a',b',c')$, then Spoiler wins 
$\game^{15}_{6\log n+28}(G,a,b,c,G',a',b',c')$.
\item
If $Q(b_1,a_1,a_2,b_2)\ne Q'(b'_1,a'_1,a'_2,b'_2)$, then Spoiler wins 
$\game^{15}_{6\log n+28}(G,b_1,a_1,a_2,b_2,G',b'_1,a'_1,a'_2,b'_2)$.
\end{bfenumerate}
\end{lemma}

\begin{proof}
{\bf 1.}
Suppose that $T(a,b,c)=0$ while $T'(a',b',c')=1$; the other case is symmetric.
The former implies that $\deg a\ge4$ and in the embedding of $G$ the vertices
$b$ and $c$ are separated by vertices $s,t\in\Gamma(a)\setminus\{b,c\}$.
Spoiler pebbles such $s$ and $t$. Let Duplicator respond with 
$s',t'\in\Gamma(a')\setminus\{b',c'\}$. Without loss of generality,
suppose that, if in the embedding of $G'$ we go around $a'$ in the order
$b'$, $c'$ and so on, then we meet $s'$ before $t'$ (otherwise just change
the notation by transposing $s$ and $t$).

Consider $X$-configurations
$C=
\begin{array}{ccccc}
u&x&y&v&w\\
s&b&t&c&a
\end{array}
$
and
$C'=
\begin{array}{ccccc}
u'&x'&y'&v'&w'\\
s'&b'&t'&c'&a'
\end{array}
$.
Here the bottom row consists of vertices and the top row of their labels.
Clearly, $C$ is collocated. Since the configuration
$\tilde C'=
\begin{array}{ccccc}
u'&x'&y'&v'&w'\\
s'&t'&b'&c'&a'
\end{array}
$
is collocated, the $C'$ is twisted.
By Lemma \ref{lem:colltwist}, Spoiler wins having made at most
$6\log n+28$ moves at total.

{\bf 2.}
Let, say, $Q(b_1,a_1,a_2,b_2)=0$ and $Q'(b'_1,a'_1,a'_2,b'_2)=1$.
To avoid considering a few trivial cases, we assume that both
$b_1,a_1,a_2,b_2$ and $b'_1,a'_1,a'_2,b'_2$ are paths.
Assume also that $T(a_2,a_1,b_2)=T(a_1,a_2,b_1)=1$ because otherwise
we are done by Item 1 of the lemma. Spoiler pebbles the vertices $c_1$ and
$c_2$ in $G$ such that
$C=
\begin{array}{cccccc}
x&y&z&u&v&w\\
c_2&b_2&a_2&b_1&c_1&a_1
\end{array}
$
is a collocated $H$-configuration. 
Denote Duplicator's responses by $c'_1$ and $c'_2$.
Unless we arrive at the conditions of Item 1, the configuration
$C'=
\begin{array}{cccccc}
x'&y'&z'&u'&v'&w'\\
c'_2&b'_2&a'_2&b'_1&c'_1&a'_1
\end{array}
$
is twisted and Spoiler wins by Lemma~\ref{lem:colltwist}.
\end{proof}

\begin{lemma}\label{lem:bb}
Suppose that $G$ and $G'$ are non-isomorphic triconnected planar graphs.
Denote $L=L_G$ and $L'=L_{G'}$. Let $m\ge W(L,L')$ and $k=\max\{m,15\}$. Then
$$
D^k(G,G')\le D^m(L,L')+6\log n+28.
$$
\end{lemma}

\begin{proof}
We have to design a strategy for Spoiler in $\game^k(G,G')$.
He emulates $\game^m(L,L')$ following an optimal strategy for this game.
His victory in $\game^m(L,L')$ means that one of the conditions of Lemma \ref{lem:aa}
is met and hence Spoiler needs $6\log n+28$ moves to win $\game^k(G,G')$.
\end{proof}

\subsection{Proof of Theorem \protect\ref{thm:dist3}}\label{ss:proofdist3}

Let $L=L_G$ and $L'=L_{G'}$. Let $R=R_G$ and $R'=R_{G'}$ (any of the two conjugated variants
can be taken).
Applying successively Lemmas \ref{lem:bb}, \ref{lem:b}, and \ref{lem:rot},
we get
\begin{eqnarray*}
D^{15}(G,G')&\le&D^{15}(L,L')+6\log n+28\\
&\le&\max\{D^5(R,R'),D^5(R^*,R')\}+8\log n+35\\
&\le&11\log n + 43.
\end{eqnarray*}

\section{Defining a triconnected planar graph}

We now prove Theorem \ref{thm:def3}. It differs from Theorem \ref{thm:dist3},
which we already proved, by allowing $G'$ to be an \emph{arbitrary} graph
non-isomorphisc to $G$. Luckily, the proof techniques we used for Theorem \ref{thm:dist3}
are still applicable. The idea is to show that for $G'$ one of two
possibilities must be the case: Either $G'$ even locally is far from being
triconnected planar and Spoiler can efficiently exploit this difference or
$G'$ is locally indistinguishable from a triconnected planar graph, in particular,
with $G'$ we can naturally associate a rotation system, and hence Spoiler
can apply the strategy of Theorem \ref{thm:dist3} designed for
triconnected planar graphs.

Let $G$ be a triconnected planar graph on $n$ vertices.
We use the tight connection between logical distinguishability 
of two structures and the \EF\/ game on these structures.
Lemma \ref{lem:colltwist} for $X$-configurations can be rephrased as follows: 
for every collocated $X$-configuration $C$ in $G$ and
every twisted $X$-configuration $T$ in a triconnected planar graph $H$
(a possibility that $H\cong G$ is not excluded),
there is a first order formula $\Phi_{C,T}(w,x,y,v,u)$ 
of quantifier depth less than $6\log n+26$
with 15 variables, of which
the variables $w,x,y,v,u$ are free, 
such that $G,C\models\Phi_{C,T}$ and $H,T\not\models\Phi_{C,T}$.
Similar formulas $\Psi_{C,T}(z,w,x,y,v,u)$ exist for $H$-configurations.

Given a collocated $X$-configuration $C$ in $G$, define $\Phi_C$ to be the
conjunction of $\Phi_{C,T}$ over all twisted configurations $T$.
A problem with this definition  is that there are infinitely many 
triconnected planar graphs $H$ and twisted $X$-configurations $T$ in them.
However, every $\Phi_{C,T}$ has quantifier depth at most $6\log n+26$
and, as well known, over a finite vocabulary there are only finitely many 
inequivalent first order formulas of a bounded quantifier depth.
If $\Phi_{C,T_1}$ and $\Phi_{C,T_2}$ are logically equivalent, then we put in
$\Phi_C$ only one of these formulas thereby making $\Phi_C$ well-defined.
Furthermore, we define $\Phi(w,x,y,v,u)$ to be the disjunction of $\Phi_C$
over all collocated $X$-configurations $C$ in $G$.
We also suppose that $\Phi$ explicitly says that $x,y,v,u$ are pairwise
distinct and all adjacent to $w$.

Similarly, for $H$-configurations we define a formula $\Psi(z,w,x,y,v,u)$
by $\Psi=\bigvee_C(\bigwedge_T\Psi_{C,T})$ and supposing also that $\Psi$
explicitly says that $z,w,x,y,v,u$ are pairwise distinct, $x,y,w\in\Gamma(z)$,
and $u,v\in\Gamma(w)$.

Notice that the order of variables we have chosen for $\Phi(w,x,y,v,u)$
plays some role. Namely, if the 5-tuple $(w,x,y,v,u)$ is a collocated 
$X$-configuration as defined in the beginning of Section \ref{s:local},
then in the embedding of $G$ the vertices $x,y,v,u$ occur around $w$
in the order as written (see Figure \ref{fig:1}).
Introduce two permulations $\sigma=(xyvu)$ and $\tau=(xu)(yv)$.
The former corresponds to the cyclic shift of the four vertices around $w$,
the latter corresponds to a reflection changing the direction around $w$.
Define
$$
\hat\Phi(w,x,y,v,u)=
\bigwedge_{i=0}^1\bigwedge_{j=0}^3\Phi(
w,\tau^i\sigma^j(x),\tau^i\sigma^j(y),\tau^i\sigma^j(v),\tau^i\sigma^j(u)
).
$$
We now make an important observation: $\hat\Phi$ has a clear geometric meaning
for 5-tuples of vertices of $G$.

\begin{lemma}
Let $a\in V(G)$ and $b_j\in\Gamma(a)$ for all $j\le4$.
In the embedding of $G$, the vertices $b_1,b_2,b_3,b_4$ occur around $a$
in the order as written if and only if
$G,a,b_1,b_2,b_3,b_4\models\hat\Phi$.
\end{lemma}

\begin{proof}
Indeed, suppose that $b_1,b_2,b_3,b_4$ is a right order around $a$.
Then the $X$-configuration 
$
C=
\begin{array}{ccccc}
x&y&v&u&w\\
b_1&b_2&b_3&b_4&a
\end{array}
$
is collocated and remains such after reassigning the labels $x,y,v,u$
with respect to the permutation $\tau^i\sigma^j$ for any $i$ and $j$.
It remains to notice that $\Phi$ is true for any collocated $X$-configuration
by construction.

For the opposite direction, suppose that $b_1,b_2,b_3,b_4$ is a wrong order around $a$.
Coherently to the previous notation, let $\sigma=(1234)$ and $\tau=(14)(23)$.
A key observation here is that, for some permutation $\pi=\tau^i\sigma^j$,
the $X$-configuration 
$
T=
\begin{array}{ccccc}
x&y&v&u&w\\
b_{\pi(1)}&b_{\pi(2)}&b_{\pi(3)}&b_{\pi(4)}&a
\end{array}
$
is twisted. By the definition of $\Phi_{C,T}$, we have
$G,a,b_{\pi(1)},b_{\pi(2)},b_{\pi(3)},b_{\pi(4)}\models\neg\Phi_{C,T}$
for every collocated $X$-configuration $C$ in $G$. It follows that
$G,a,b_{\pi(1)},b_{\pi(2)},b_{\pi(3)},b_{\pi(4)}\models\neg\Phi_{C}$
for every $C$ and hence $G,a,b_{\pi(1)},b_{\pi(2)},b_{\pi(3)},b_{\pi(4)}\models\neg\Phi$.
Equivalently, we have 
$G,a,b_1,b_2,b_3,b_4\models\neg\Phi(w,\pi^{-1}(x),\pi^{-1}(y),\pi^{-1}(v),\pi^{-1}(u))$,
where $\pi^{-1}=\sigma^{4-j}\tau^{2-i}$.
Thus, $G,a,b_1,b_2,b_3,b_4\models\neg\hat\Phi(w,x,y,v,u)$, as required.
\end{proof}

Define a first order statement
\begin{eqnarray*}
&&
A_G=\forall x,y_1,y_2,y_3,y_4\Bigg(
\bigwedge_{i=1}^4y_i\sim x\wedge\bigwedge_{i\ne j}\neg(y_i=y_j)\to\\
&&\bigg(
\hat\Phi(x,y_1,y_2,y_3,y_4)\vee
\bigvee_\rho\hat\Phi(x,y_{\rho(1)},y_{\rho(2)},y_{\rho(3)},y_{\rho(4)})
\bigg)
\wedge
\\&&
\bigg(
\hat\Phi(x,y_1,y_2,y_3,y_4)\to
\bigwedge_\rho\neg\hat\Phi(x,y_{\rho(1)},y_{\rho(2)},y_{\rho(3)},y_{\rho(4)})
\bigg)
\Bigg),
\end{eqnarray*}
where $\rho$ ranges over all transpositions of two elements in $\{1,2,3,4\}$.
Note that $A_G$ has quantifier depth at most $6\log n+31$.
This sentence has a clear geometric meaning and is true on~$G$.

Suppose now that $G'$ is an arbitrary graph non-isomorphic to $G$.
We have to bound $D^{15}(G,G')$ from above. 
We assume that $G'$ is connected and has minimum degree at least 3;
otherwise Spoiler wins fast.
If $G'\not\models A_G$,
then $G$ and $G'$ are distinguished by $A_G$ and hence $D^{15}(G,G')\le 6\log n+31$.

Suppose that $G'\models A_G$. The $A_G$ ensures that, for every vertex $a$ in $G'$
and $b_1,b_2,b_3,b_4\in\Gamma(a)$, we have a unique (up to shifting and redirecting,
i.e., up to a permutation $\tau^{i}\sigma^{j}$)
circular ordering of $b_1,b_2,b_3,b_4$ satisfying $\hat\Phi(x,y_1,y_2,y_3,y_4)$.
We use it to associate with $G'$ a layout system $L'=\langle G',T',Q'\rangle$
(as if this ordering corresponds to some embedding of~$G'$).
Given $a\in V(G')$ of degree at least 4, we first want to define pairs
$b,c\in\Gamma(a)$ such that $b$ and $c$ are neighboring in this ``pseudo-embedding''
of $G'$. 

We let $N(a,b,c)=\neg\exists s,t\hat\Phi(a,b,s,c,t)$. Consider a first order sentence
$$
B_G=\forall a,b\Big(
\deg a\ge4\wedge b\sim a\to\exists_{=2}c\,N(a,b,c)
\Big)
$$
(written with harmless shorthands).
This sentence has a clear geometric meaning and is true on $G$.
If $G'\not\models B_G$,
then $G$ and $G'$ are distinguished by $B_G$ and we are done.

Suppose that $G'\models B_G$. We are now able to define a ternary relation $T'$
on $V(G')$.
Suppose that $b',c'\in\Gamma(a')$ and $b'\ne c'$. If $\deg a'=3$, we set 
$T'(a',b',c')=1$. Let $\deg a'\ge4$. In this case we set $T'(a',b',c')=1$
iff $N(a',b',c')$ is true.

The $B_G$ ensures that, for every $a'$, $T'_{a'}$ is a union of cycles.
If $T'_{a'}$ is disconnected for some $a'$, Spoiler wins fast. He first pebbles the
$a'$. Denote Duplicator's response in $G$ by $a$. Spoiler restricts further play
to $\Gamma(a)$ and $\Gamma(a')$ and follows his winning strategy in the game
on graphs $(T_G)_a$ and $T'_{a'}$, one of which is connected and the other is not.
Spoiler's win in this game entails disagreement $N(a,b,c)\ne N(a',b',c')$
for some pebbled $b,c$ in $G$ and the corresponding $b',c'$ in $G'$.
In the next two moves Spoiler forces disagreement between the truth values
of $\hat\Phi$ on some 5-tuples and wins in $6\log n+26$ extra moves.

Suppose hence that $T'_{a'}$ is connected for every $a'$, i.e, is a cycle on $\Gamma(a')$.
Similarly to the above, we can use the formula $\Psi$ to construct a sentence
$\Lambda_G$ of quantifier depth at most $6\log n+32$ providing us with
the following dichotomy. If $G'\not\models\Lambda_G$, the $G$ and $G'$ are
distinguished by $\Lambda_G$ and we are done. Otherwise $\Psi$ in a natural way
determines a quaternary relation $Q'$ such that $L'=\langle G',T',Q'\rangle$
is a layout system.

We have to consider the latter possibility. In its turn, it splits into two cases.
If $L'=\lay(R')$ for no rotation system $R'$, this means that, if we fix a triple
$a'_1,b'_1,c'_1$ with $T'(a'_1,b'_1,c'_1)=1$ and set $T'_{R'}(a'_1,b'_1,c'_1)=1$, 
then there are a triple $a'_2,b'_2,c'_2$
and two $a'_1$-$a'_2$-paths $P_1$ and $P_2$ such that propagation of the truth value
of $T'_{R'}(a'_1,b'_1,c'_1)$ along $P_1$ and $P_2$ gives different results, say,
$T'_{R'}(a'_2,b'_2,c'_2)=1$ for $P_1$ and $T'_{R'}(a'_2,c'_2,b'_2)=1$ for $P_2$.
Spoiler pebbles $a'_1,b'_1,c'_1,a'_2,b'_2,c'_2$. Let Duplicator
respond with $a_1,b_1,c_1,a_2,b_2,c_2$ in $G$. Suppose that $T_R(a_1,b_1,c_1)=1$
for $R\in\{R_G,R^*_G\}$. Spoiler wins similarly to the proof of Lemma \ref{lem:a},
using $P_1$ if $T_R(a_2,c_2,b_2)=1$ and $P_2$ if $T_R(a_2,b_2,c_2)=1$.

If $L'=\lay(R')$ for some rotation system $R'$, then Spoiler plays as if $G'$ was
a triconnected planar graph. Namely, he follows the strategy of Section \ref{s:global}
using $L'$ for $L_{G'}$ and $R'$ for $R_{G'}$. Spoiler's win in this simulations means
that he forces pebbling some tuples of vertices in $G$ and $G'$ on which the formula
$\Phi$ or the formula $\Psi$ disagree, and hence logarithmically many extra moves
suffice for Spoiler to have a win in $\game^{15}(G,G')$.

\subsection*{Acknowledgment}
I acknowledge valuable discussions with Martin Grohe on the topic
and am grateful to Hans-J\"urgen Pr\"omel for his kind hospitality
during my two-year research stay at the Humboldt-University of Berlin.

\end{document}